%% file: arXiv_Report_TCP_V2X.tex
\documentclass[11pt]{article}
 
\usepackage[margin=1in]{geometry} 
 \usepackage{amsmath}
\usepackage{amssymb}
\usepackage{multirow}
\usepackage{floatrow}
\usepackage{threeparttable}
\usepackage[english]{babel}
\usepackage[T1]{fontenc}
\usepackage[utf8]{inputenc}
\usepackage{float}
\usepackage{verbatim}
\usepackage{tabularx}
\usepackage{alltt}
\usepackage{cite}
\usepackage[table,xcdraw]{xcolor}
\usepackage{algorithm}
\usepackage{algpseudocode} 
\usepackage{ifthen}   

   \usepackage[nottoc]{tocbibind}
   \usepackage{multirow}
   \usepackage{booktabs}
   \usepackage{graphicx}
   \usepackage{subcaption}
   \usepackage{epstopdf}
   \usepackage{array, blindtext}
   \usepackage{wrapfig}
\usepackage{pdfpages}
\usepackage{array}
\usepackage[font=small,labelfont=bf]{caption}
\usepackage{lipsum}
\usepackage{titlesec}  %change space above chapter sections

\usepackage{tikz}
\usepackage{pgfplots}
\pgfplotsset{compat=newest} 
\pgfplotsset{plot coordinates/math parser=false} 
\newlength\fheight
\newlength\fwidth
\usetikzlibrary{patterns,decorations.pathreplacing,backgrounds,calc}
\definecolor{SchoolColor}{RGB}{0.71, 0, 0.106}%181,0,27} unipd red
\definecolor{chaptergrey}{rgb}{0.61, 0, 0.09} % dialed back a little
\definecolor{midgrey}{rgb}{0.4, 0.4, 0.4}
\definecolor{chaptergreen}{rgb}{0.09, 0.612, 0}
\definecolor{chapterpurple}{rgb}{0.522, 0, 0.612}
\definecolor{chapterlightgreen}{rgb}{0, 0.612, 0.522}

\newcommand{\R}{R_{\rm comm}}
\newcommand{\T}{T_{\rm RTO}}

\newcommand{\PNL}{P_{\rm NL}}
\usepackage[english]{isodate}

\usepackage{titlesec}
\titleformat{\section}[block]
  {\fontsize{12}{15}\bfseries\filcenter}
  {\thesection}
  {1em}
  {\MakeUppercase}
\titleformat{\suINection}[hang]
  {\fontsize{12}{15}\bfseries\sffamily}
  {\thesuINection}
  {1em}
  {}
\usepackage{fancyhdr}

\fancypagestyle{style1}{
\fancyfoot[R]{\thepage}
\fancyhead[R]{}

}
\usepackage{datetime}

\usetikzlibrary{automata,positioning}

\newif\iftikz
\tikztrue
 
\begin{document} 
 
% --------------------------------------------------------------
%                         Start here
% --------------------------------------------------------------
\title{\textbf{Technical Report} \\\vspace{5mm} Millimeter Wave Communication in Vehicular Networks: Coverage and Connectivity Analysis}
\author{\textbf{Marco Giordani} \and \textbf{Andrea Zanella} \and \textbf{Michele Zorzi}\\ %replace with your name
\and \footnotesize E-mail: {\{giordani, zanella, zorzi\}@dei.unipd.it}} %if necessary, replace with your course title
\date{\printdayoff \today}
\maketitle
\thispagestyle{style1}
%\small

In this technical report (TR), we will report the mathematical model we developed to carry out the preliminary coverage and connectivity analysis in mmWave-based vehicular networks, proposed in our work \cite{MOCAST17}. The purpose is to exemplify some of the complex and interesting tradeoffs that have to be considered when designing solutions for mmWave automotive scenarios. 

The rest of TR is organized as follows. 
In Section \ref{sec:scenario}, we describe the scenario used to carry out our simulation results, presenting the main setting parameters and the implemented mmWave channel model. In Section \ref{sec:model} we report the mathematical model used to  develop our connectivity and coverage analysis. Finally, in Section \ref{sec:analysis}, we show our main findings in terms of~throughput.

\section{Simulation Settings}
\label{sec:scenario}

We  consider a simple but representative V2I scenario, where a single Automotive Node (AN, i.e., a car) moves along a road at constant speed $V$ and Infrastructure Nodes (INs, i.e., static mmWave Base Stations) are randomly distributed according to a Poisson Point Process (PPP) of parameter $\rho$ nodes/km, so that the distance $d$ between consecutive nodes is an exponential random variable of mean $\mathbb{E}[d]= 1/\rho$ km (see, e.g., \cite{leutzbach1988introduction}).
%According to \cite{rudack2002dynamics}, AN move at random speeds $V$, which are generally assumed to be normal random variables with mean $\mu$ and variance $\sigma^2$. 
%Roadside units, also called infrastructure nodes (INs), like mmWave base station (BS), are instead static. Nonetheless, we still assumed 
%When considering a V2I configuration, we state that the relative speed between vehicle and infrastructure is~$\Delta V = V$. 

\subsection{Millimeter Wave Channel Model}
As assessed in \cite{MOCAST17}, in order to overcome the increased isotropic
path loss experienced at higher frequencies, next-generation mmWave automotive communication
must provide mechanisms by which the vehicles and the infrastructures  determine suitable  directions
of transmission for spreading around their sensors information, thus exploiting beamforming (BF) gain at both the transmitter and the receiver side.
To provide a realistic assessment of mmWave micro and picocellular networks in a dense urban deployment, we considered the channel model obtained from recent real-world measurements at $28$ GHz in New York City. Further details on the channel model and its parameters can be found in \cite{Mustafa}.\footnote{As pointed out in \cite{MOCAST17}, available measurements at mmWaves in the V2X context are still very limited, and realistic scenarios are indeed hard to simulate.
Moreover, current models for mmWave cellular systems (e.g., \cite{Mustafa}) present many limitations for their applicability to a V2X context, due to the more challenging propagation characteristics of highly mobile vehicular nodes. Simulating more realistic scenarios for further validating the presented results, like considering channel models specifically tailored to a V2X context, is of great interest an will be part of our future analysis.}

The link budget for the mmWave propagation channel is defined as:
\begin{equation}
P_{RX} = P_{TX} + G_{BF} - PL - \xi
\label{eq:lbud}
\end{equation}
where $P_{RX}$ is the total received power expressed in dBm, $P_{TX}$ is the transmit power, $G_{BF}$ is the gain obtained using BF techniques, $PL$ represents the pathloss in dB and ${\xi\sim N(0,\sigma^2)}$ is the shadowing in dB, whose parameter $\sigma^2$ comes from the measurements in \cite{Mustafa}.

Based on the real-environment measurements of \cite{Mustafa}, the pathloss can be modeled through three different states: Line-of-Sight (LoS), Non-Line-Of-Sight (NLoS) and outage. Based on the distance $d$ between the transmitter and the receiver, the probability to be in one of the states ($P_{\text{LoS}}$, $P_{\text{NLoS}}$, $P_{\text{out}}$) is computed by:
\begin{equation}
\begin{split}
P_{\text{out}}(d) &= \max(0,1-e^{-a_{\text{out}}d+b_{\text{out}}})\\
P_{\text{LoS}}(d) &= (1-P_{\text{out}}(d) )e^{-a_{\text{LoS}}d}\\
P_{\text{NLoS}}(d) &= 1- P_{\text{out}}(d) - P_{\text{LoS}}(d)
\end{split}
\end{equation}
where parameters ${a_{\text{out}} = 0.0334}$  m$^{-1}$, ${b_{\text{out}} = 5.2}$  and ${a_{\text{LoS}} = 0.0149}$  m$^{-1}$ have been obtained in \cite{Mustafa} for a carrier frequency of $28$ GHz. The pathloss is finally obtained by:
\begin{equation}
PL(d)[dB] = \alpha + \beta 10 \log_{10}(d) 
\end{equation} 
where $d$ is the distance between receiver and transmitter, and the value of the parameters $\alpha$ and $\beta$ are given in \cite{Mustafa}.

\pagestyle{fancy}

To generate random realizations of  the large-scale parameters, the mmWave channel is defined as a combination of a random number $K\sim \max\{\text{Poisson}(\lambda),1 \}$  of path clusters, for which the parameter $\lambda$ can be found in \cite{Mustafa}, each  corresponding to a macro-level scattering path. Each cluster is further composed of several subpaths $L_k \sim U[1,10]$. 

The time-varying  channel matrix is described as follows:
\begin{equation}
\textbf{H}(t,f) = \frac{1}{\sqrt{L}} \sum_{k=1}^{K} \sum_{l=1}^{L_k} g_{kl}(t,f) \textbf{u}_{rx}(\theta_{kl}^{rx},\phi_{kl}^{rx})\textbf{u}^*_{tx}(\theta_{kl}^{tx},\phi_{kl}^{tx})
\label{eq:H}
\end{equation}
where $g_{kl}(t,f)$ refers to the small-scale fading over time and frequency on the $l^{\text{th}}$ subpath of the $k^{\text{th}}$ cluster and $\textbf{u}_{rx}(\cdot)$, $\textbf{u}_{tx}(\cdot)$ are the spatial signatures for the receiver and transmitter antenna arrays and are functions of the central azimuth (horizontal) and elevation (vertical) Angle of Arrival (AoA) and Angle of Departure (AoD),  respectively $\theta_{kl}^{rx}$, $\phi_{kl}^{rx}$, $\theta_{kl}^{tx}$, $\phi_{kl}^{tx}$ \footnote{Such angles can be generated as wrapped Gaussian around the cluster central angles with standard deviation given by the rms angular spread for the cluster given in \cite{Mustafa}.}.

The small-scale fading in Equation \eqref{eq:H} describes the rapid fluctuations of the amplitude of a radio signal over a short period of time or travel distance. It is generated based on the number of clusters, the number of subpaths in each cluster, the Doppler shift, the power spread, and the delay spread, as:

\begin{equation}
g_{kl}(t,f)=\sqrt{P_{lk}}e^{2\pi i f_d \cos(\omega_{kl})t-2\pi i \tau_{kl}f},
\label{eq:small_scale_fading}
\end{equation}
where:
\begin{itemize}
\item  $P_{lk}$ is the power spread of  subpath $l$ in  cluster $k$, as defined in \cite{Mustafa};
\item $f_{d}$ is the maximum Doppler shift and is related to the  user speed ($v$) and to the carrier frequency $f$ as $f_d = f v / c$, where $c$ is the speed of light;
\item $\omega_{kl}$  is the angle of arrival of subpath $l$ in cluster $k$ with respect
 to the direction of motion;
\item $\tau_{kl}$ gives the delay spread of subpath $l$ in  cluster $k$;
\item $f$ is the carrier frequency.
\end{itemize}

Due to the high pathloss experienced at mmWaves, multiple antenna elements with beamforming are essential to provide an acceptable  communication range. The BF gain parameter in \eqref{eq:lbud} from transmitter $i$ to receiver $j$ is thus given by:
\begin{equation}
G_{BF}(t,f)_{ij} = |\textbf{w}^*_{rx_{ij}}\textbf{H}(t,f)_{ij}\textbf{w}_{tx_{ij}}|^2
\label{beamforming_gain}
\end{equation}
where $\textbf{H}(t,f)_{ij}$ is the channel matrix of the $ij^{th}$ link, $\textbf{w}_{tx_{ij}}\in \mathbb{C}^{n_{\mathbb{T}x}}$ is the BF vector of transmitter $i$ when transmitting to receiver $j$, and $\textbf{w}_{rx_{ij}}\in \mathbb{C}^{n_{\mathbb{R}x}}$ is the BF vector of receiver $j$ when receiving from transmitter $i$.  Both vectors are complex, with length equal to the number of antenna elements in the array, and are chosen according to the specific direction that links BS and UE.\\

%
%
%Statistical models are derived for key channel parameters, including: (i) a distance-based pathloss, which models line-of-sight (LOS), non-line-of-sight (NLOS) and outage conditions; (ii) spatial clusters, described by central azimuth and elevation angles, fractions of power and angular beamspreads;  (iii) a small-scale fading model, where each of the path clusters is synthesized with a large number of subpaths, each  having its own peculiarities on horizontal and vertical angles (generated around the cluster central angles). Further details on the channel model and its parameters can be found in \cite{Mustafa,ns3_nokia,rappaport_channel_model}. 

The channel quality is measured in terms of Signal-to-Interference-plus-Noise-Ratio (SINR). By referring to the mmWave statistical channel described above, the SINR between a transmitted $\rm j$ and a test RX can computed in the following way:
\begin{equation}
\Gamma = \text{SINR}_{\rm j,RX} = \frac{\frac{P_{\rm TX}}{PL_{\rm j,UE}}G_{\rm j,RX}}{\sum_{\rm k\neq j}\frac{P_{\rm TX}}{PL_{\rm k,RX}}G_{\rm k,RX}+W_{\rm tot}\times N_0}
\label{eq:SINR}
\end{equation}
where $G_{\rm i,RX}$ and $PL_{\rm i,RX}$ are the BF gain and the pathloss obtained between transmitter $ i$ and the test RX, respectively, and $ W_{\rm tot}\times N_0$ is the thermal noise.

\subsection{System Model}

We say that an AN is within coverage $(\R)$ of a certain IN if, assuming perfect beam alignment, the SINR is the best possible for the AN, and it exceeds a minimum threshold, which we set to $\Gamma_0=-5$~dB. Due to the stochastic nature of the signal propagation and of the interference, the coverage range of an IN is a random variable, whose exact characterization is still unknown but clearly depends on a number of factors, such as beamwidth,  propagation environment, and level of interference that, in turn, depends on the spatial density of the nodes.\\ 

\begin{figure}[!b]
\centering
\iftikz
	\setlength{\belowcaptionskip}{0cm}
	\setlength\fwidth{0.351\columnwidth}
	\setlength\fheight{0.306\columnwidth}
	\input{./Figures/Rcomm_rho.tex}
	\else
	\includegraphics[width=0.4\textwidth]{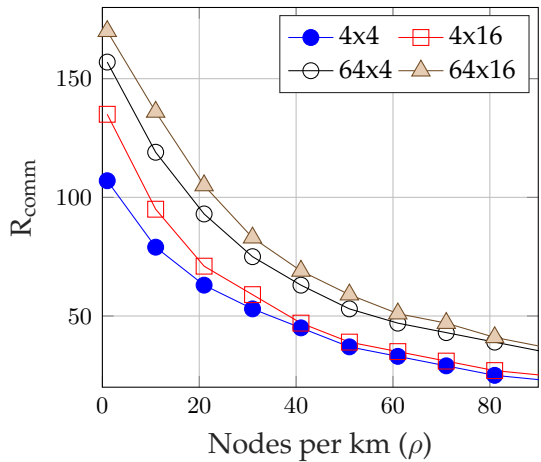}
	\fi
\caption{Average coverage range $\R$ (for different MIMO configurations) versus the nodes density $\rho$.}
\label{fig:Rcomm_rho}
\end{figure}

To gain some insights on these complex relationships, we  performed a number of simulations and evaluated the mean coverage range $\R$ when varying the node density and the antenna configuration of the nodes.
Tab.~\ref{tab:params} collects the main simulation parameters, which are based on realistic system design considerations. 
A set of two dimensional antenna arrays is used at both the INs and the ANs. INs are equipped with a Uniform Planar Array (UPA) of $2 \times 2 $ or $8 \times 8$ elements, while the ANs exploit an array of $2 \times 2 $ or $4 \times 4$ antennas. The spacing of the elements is set to $\lambda/2$, where $\lambda$ is the wavelength.  
In general,  as depicted in Fig.~\ref{fig:Rcomm_rho}, $\R$ increases with the number of antennas, thanks to the narrower beams that can be realized, which increase the beamforming gain. 
On the other hand, $\R$ decreases as the density of nodes increases, because of the larger amount of interference received at the AN from the INs due to their reduced distance.

\begin{table}[!t]
\centering
\renewcommand{\arraystretch}{1.2}% Tighter
\resizebox{0.65\columnwidth}{!}%
{
\begin{tabular}{|c|c|c|}
\hline
\textbf{Parameter} & \textbf{Value} & \textbf{Description}\\
\toprule
\hline
 $W_{\rm tot}$ & $1$ GHz & Total system bandwidth\\
\hline
DL  $P_{\rm TX}$ & $30$ dBm & Transmission power \\
\hline
 NF  & $5$ dB & Noise figure \\
\hline
$f_{\rm c}$ & $28$ GHz & Carrier frequency \\
\hline
$\tau$ & $ -5$ dB &  Minimum SINR threshold \\
\hline
$N_{\rm BS}$ & \{4, 64\} & BS MIMO array size\\
\hline
$N_{\rm veh}$ & \{4, 16\} & Vehicular MIMO array size\\
\hline
$\T$ & \{0.025, 0.1, 0.2, 0.5, 1\} s  & Slot duration \\
\hline
$ V$ & \{10, 20, 30, 100, 130\} km/h & Vehicle speed \\
\hline
$\R$ & Varied & Communication radius\\
\hline
\end{tabular}
}
\caption{Main simulation parameters.}
\label{tab:params}
\end{table}

As we pointed out in \cite{MOCAST17},  a directional beam pair needs to be determined to  enable the transmission between two ANs, thus beam tracking heavily affects the connectivity performance of a V2X mmWave scenario.
In this  analysis, according to the procedure described in \cite{giordani2016uplink,giordani_MedHoc2016}, we assume that  measurement reports are periodically exchanged among the  nodes so that, at the beginning of every  slot of duration $\T$, ANs and INs identify the best directions for their respective  beams. 
Such configuration is kept fixed for the whole slot duration, during which nodes may lose the alignment due to the AN mobility. In  case the connectivity is lost during a slot, it can only be recovered at the beginning of the subsequent slot, when the beam tracking procedure is performed again. 
%\footnote{As part of our future work, advanced beam tracking procedures will be investigated, according to the proposals that we presented in Section \ref{sec:PHY-MAC_solutions} and aiming at overcoming the well-known mmWave limitations for the design of a proper MAC layer of  V2X automotive systems.}
It is hence of interest to evaluate the AN connectivity, i.e., the fraction of slots in which the AN and the IN remain connected (since the AN is in the  IN's coverage range), as a function of the following parameters: (i) the MIMO configuration; (ii) the vehicle speed $V$; (iii) the slot duration $T_{\rm RTO}$; (iv) the node density $\rho$.\\

%\begin{figure}
%\begin{tikzpicture}
%\node[state, fill=brown!40]                               (0) {C};
%\node[state,right=of 0]                    (1) {I};
%\draw[
%    >=latex,
%   every node/.style={above,midway},% either
%    auto=right,                      % or
%    loop above/.style={out=75,in=105,loop},
%    every loop,
%    ]
%     (1)   edge[loop above] node {$p_{CC}$}   (1)
%           edge [bend right]           node {$p_{IC}$}   (0)
%     (0)   edge[loop above, draw = brown!40, line width = 1 mm] node {$p_{CC}$}   (0)
%      (0)  edge  [bend right] node   {$p_{CI}$}   (1);
%\end{tikzpicture}
%\caption{Prova.}
%\end{figure}

At the beginning of a time slot, the AN can be either  in a connected (C) state, if it is within the coverage range of an IN, or in an idle (I) state, if there are no INs within a distance~$\R$. 
Starting from state C, the AN can either maintain connectivity to the serving IN for the whole slot duration, or lose the beam alignment and get disconnected.  
Starting from state I, instead, the AN can either remain out-of-range for the whole slot of duration $\T$, or enter the coverage range of a new IN within $\T$ (\emph{catch-up}). Even in this second case, however, the connection to the IN will be established only at the beginning of the following slot, when the beam alignment procedure will be performed. 
Therefore,  preservation of the connectivity during a slot requires that the AN is within the coverage range of the IN at the beginning of the slot and does not lose beam alignment in the slot period $\T$. In this case, the vehicle can potentially send data with a rate $R(d)$ that depends on the distance $d$ to the serving IN. 

\section{Coverage and Connectivity Analysis}
\label{sec:model}

In order to determine the average throughput  a vehicular node will experience in the considered simple automotive scenario, as a function of several V2X parameters, we first need to  compute the average portion of slot in which the VN is both within the coverage of an infrastructure node and properly aligned. To do so, in this section we will: (i) evaluate the probability for the VN to be within $\R$, at the beginning of the slot; (ii) evaluate the probability for the VN not to misalign within the slot; (iii) finally evaluate the mean communication duration.
 
\subsection{Probability of Starting the Communication}

\begin{figure}[!h]
\centering
	\includegraphics[width=0.7\textwidth]{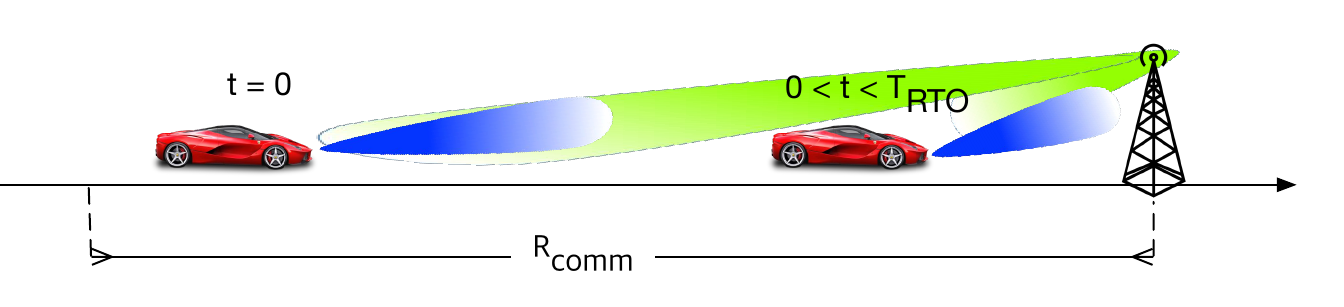}
\caption{Scenario in which the VN is both within the coverage range $\R$ of its serving IN and properly aligned for the whole slot of duration $\T$.}
\label{fig:CC}
\end{figure}

The communication can start, within the slot of duration $T_{\rm RTO}$, only if the vehicle is already within the coverage range $R_{\rm comm}$, with probability $\mathbb{P}_D (R_{\rm comm})$:

\begin{equation}
P_{\rm start} = \mathbb{P}[d \leq 2 R_{\rm comm}] = 1-e^{-2\rho R_{\rm comm}}
\end{equation}

\begin{figure}[!t]
     \centering
\iftikz
	\setlength{\belowcaptionskip}{0cm}
	\setlength\fwidth{0.7\columnwidth}
	\setlength\fheight{0.3\columnwidth}
	\input{./Figures/P_start.tex}
	\else
	\includegraphics[width=0.4\textwidth]{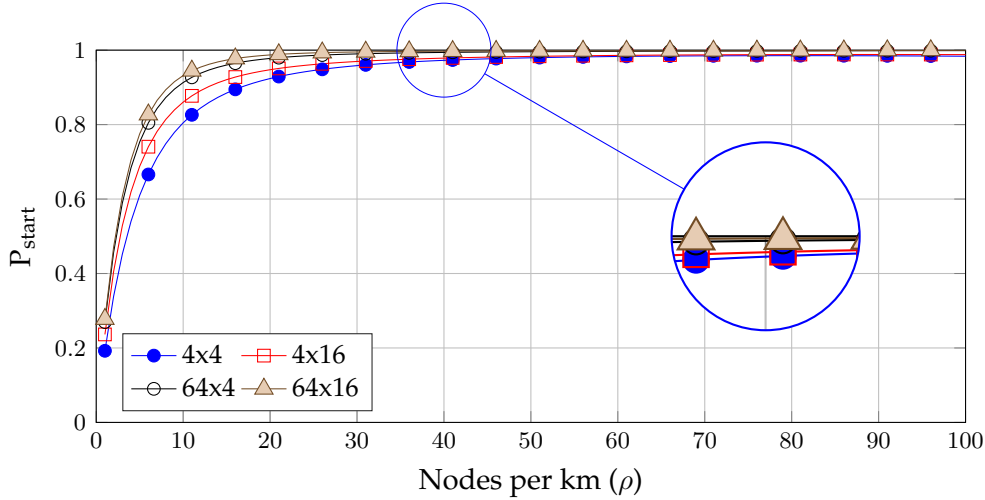}
	\fi
\caption{Probability of starting the communication vs. nodes density $\rho$, when VN moves at constant speed $V=90$ km/s and $\T=200$ ms.}
\label{fig:P_start}
\end{figure}

%\begin{figure}[!t]
%     \centering
% \begin{subfigure}[t!]{0.45\columnwidth}
%     		\setlength{\belowcaptionskip}{0cm}
%\centering
%\iftikz
%	\setlength{\belowcaptionskip}{0cm}
%	\setlength\fwidth{0.8\columnwidth}
%	\setlength\fheight{0.8\columnwidth}
%	\input{./Figures/P_start.tex}
%	\else
%	\includegraphics[width=0.4\textwidth]{Figures/P_start}
%	\fi
%\caption{$P_{\rm start}$ vs. nodes density.}
%\end{subfigure}\hfill 
%\begin{subfigure}[t!]{0.45\columnwidth}
%\setlength{\belowcaptionskip}{0cm}
%\centering
%\iftikz
%	\setlength{\belowcaptionskip}{0cm}
%	\setlength\fwidth{0.8\columnwidth}
%	\setlength\fheight{0.8\columnwidth}
%	\input{./Figures/P_start_2.tex}
%	\else
%	\includegraphics[width=0.4\textwidth]{Figures/P_start}
%	\fi
%\caption{$P_{\rm start}$ vs. nodes density (zoom for sparse infrastructure nodes).}
%\end{subfigure}
%\caption{Probability of starting the communication.}
%\label{fig:P_start}
%\end{figure}

\noindent From the results in Figure \ref{fig:P_start}, we deduce that:

\begin{itemize}
\item $P_{\rm start}$ increases with the IN density $\rho$, since the mean distance $\mathbb{E}[d] = 1/\rho$ from the IN decreases, so it's more likely for the AN to fall within the coverage range of the IN. 
On the other hand, although the communication range $R_{\rm comm}$ is reduced when considering denser networks, due to the increased interference perceived by the vehicular node, $P_{\rm start}$ still  increases, making the reduction of the distance \textbf{dominant} to the increased interference\footnote{We observe that the increasing behavior of $P_{\rm start}$ saturates when INs are particularly dense.}.

\item $P_{\rm start} $ increases when increasing the MIMO order, that is when packing more antenna elements in the MIMO array. In fact, keeping $\rho$ fixed, beams are narrower, interference is reduced, the achieved BF gain is higher, and the increased $R_{\rm comm} $ makes the discoverable range of the IN increased as well.
\end{itemize}

\subsection{Probability of NOT Leaving the Communication}

Constraining on the probability of having started the communication within the slot of duration $\T$ , the vehicle loses its ability to communicate with the IN with probability $1-P_{\rm NL}$, where $\PNL$ is defined as:
\begin{equation}
P_{\rm NL} = P_{\rm start}\cdot\mathbb{P}(T_{\rm L} > T_{\rm RTO}).
\label{eq:P_NL}
\end{equation}

\begin{figure}[!h]
\centering
	\includegraphics[width=0.7\textwidth]{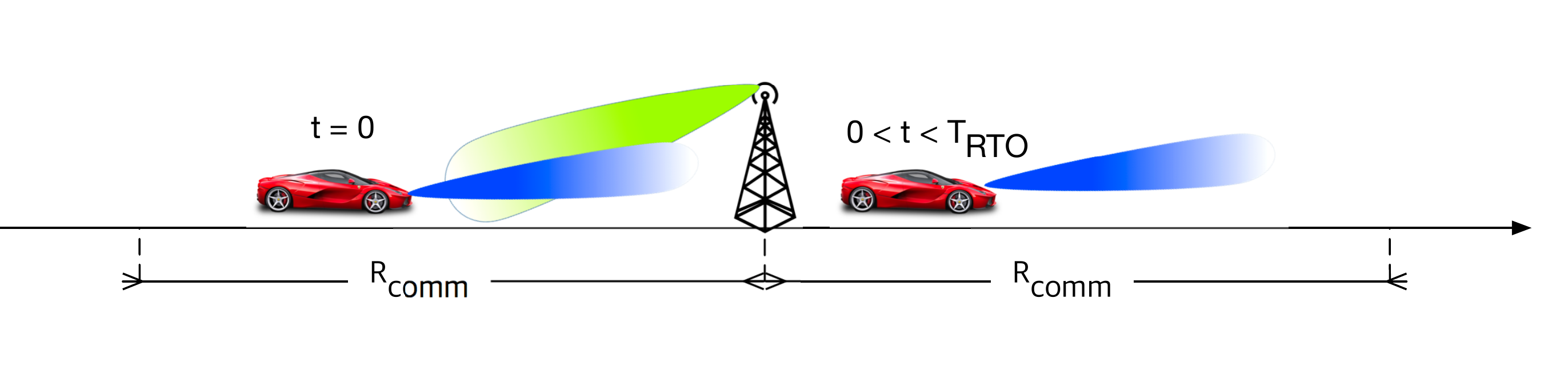}
\caption{At the beginning of the slot, the AN is connected to the IN, steering the beam to an angle of 0$^\circ$. When moving at constant speed $V$ during the slot, the AN overtakes the IN (now on its left). Since the beam direction cannot be updated during the slot, the link between the AN  and IN will be lost till the beginning of the next slot. }
\label{fig:overtake}
\end{figure}

\noindent If the  vehicle was already in the communication range at the beginning of the slot, it does not leave the communication with probability $\mathbb{P}(T_{\rm L} > T_{\rm RTO})$, that is if it covers a distance smaller than $d|d<\R$, within $\T$, moving at relative speed $ V>0$: 

\begin{align}
& \mathbb{P}(T_{\rm L} > T_{\rm RTO}) = \\ \notag
& =1-\mathbb{P}(\text{Overtaking the IN within } \T | \text{vehicle is connected}) = \\ \notag
&=  \mathbb{P}\Big[ \dfrac{d}{V} > \T | d<\R \Big] = \frac{e^{-\rho\T V}-e^{-\rho\R}}{1-e^{-\rho\R}} 
\end{align}

On the other hand, if the AN overtakes the IN (as in Figure \ref{fig:overtake}) during the slot, although theoretically being under the coverage of the infrastructure, it  needs to adapt its beam orientation to be perfectly aligned; however, this operation can be triggered only at the beginning of the subsequent slot, making the AN misaligned and thus disconnected for the whole remaining slot period.

\begin{figure}[t!]
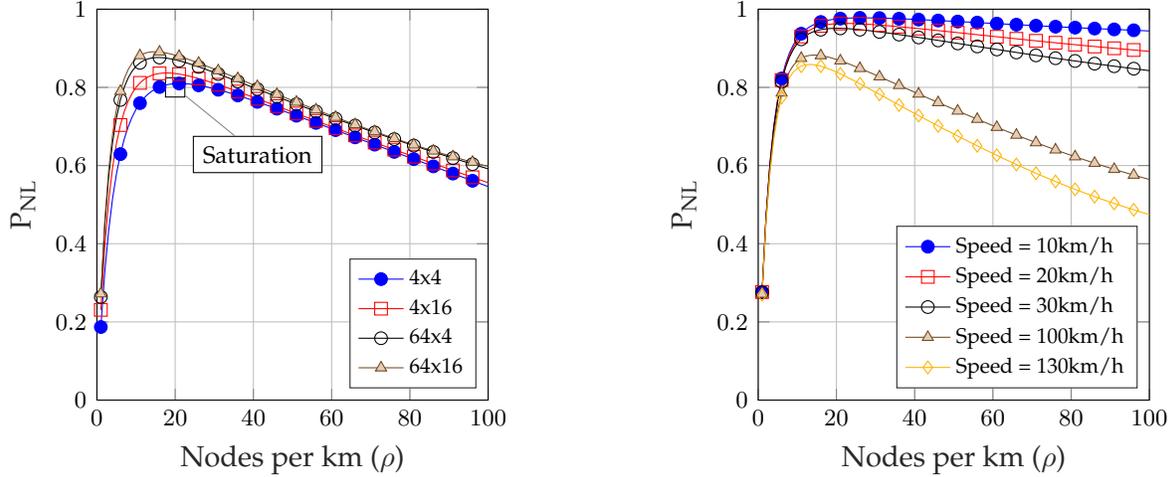
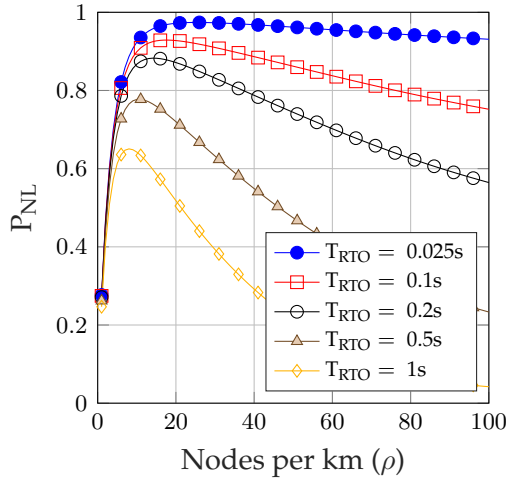

\centering
     \begin{subfigure}[t!]{0.45\textwidth}
     \centering
     		\setlength{\belowcaptionskip}{0cm}
      \iftikz
	\setlength{\belowcaptionskip}{0cm}
	\setlength\fwidth{0.7\columnwidth}
	\setlength\fheight{0.7\columnwidth}
	\input{./Figures/P_NL_MIMO.tex}
	\else
   \includegraphics[width=\textwidth]{Figures/P_NL_MIMO.eps}
    \fi
        \caption{$P_{\rm NL}$ with $V=90$ km/s and $\T = 200$ ms, for different MIMO configurations.}
   \end{subfigure} \hfill
             \begin{subfigure}[t!]{0.45\textwidth}
             		\setlength{\belowcaptionskip}{0cm}
             \iftikz
	\setlength{\belowcaptionskip}{0cm}
	\setlength\fwidth{0.7\columnwidth}
	\setlength\fheight{0.7\columnwidth}
	\input{./Figures/P_NL_V.tex}
	\else
   \includegraphics[width=\textwidth]{Figures/P_NL_V.eps}\fi
    \caption{$P_{\rm NL}$ with $\T = 200$ ms and MIMO configuration $64\times 16$, for different speeds.}
      \end{subfigure} \qquad
      \begin{subfigure}[t!]{0.45\textwidth}
      		\setlength{\belowcaptionskip}{0cm}
  \iftikz
	\setlength{\belowcaptionskip}{0cm}
	\setlength\fwidth{0.7\columnwidth}
	\setlength\fheight{0.7\columnwidth}
	\input{./Figures/P_NL_T.tex}
	\else
   \includegraphics[width=\textwidth]{Figures/P_NL_T.eps} \fi
    \caption{$P_{\rm NL}$ with $V=90$ km/h and MIMO configuration $64\times 16$, for different slot durations.}
      \end{subfigure}
\caption{Probability of not leaving the communication ($P_{\rm NL}$) within a time slot of duration $\T$ when varying the nodes spatial density $\rho$.}
\label{fig:P_NL}
\end{figure}

In Figure \ref{fig:P_NL}, we plot the probability of \textbf{not} leaving the communication ($P_{\rm NL}$) and we state that:
\begin{itemize}
\item $P_{\rm NL}$ increases with $\rho$, for sparse networks (\textbf{$\rho$ small}). On one hand, $P_{\rm start}$ increases but, on the other hand, the AN is closer and closer to the IN (the mean distance $\mathbb{E}[d] = 1/\rho$ from the IN decreases) and, within the same time slot $T_{\rm RTO}$, it is more and more likely for the AN to overtake the IN and being misaligned. 
However, when the infrastructure nodes are quite scattered,  the mean distance $\mathbb{E}[d] = 1/\rho$ is relatively large and so the increasing behavior of $P_{\rm start}$ in Eq. \eqref{eq:P_NL} is \textbf{dominant}.
 \item $P_{\rm NL}$ decreases with increasing values of $\rho$, for dense networks (\textbf{$\rho$ large}). In fact, $P_{\rm start}$ has reached a quasi-steady state (see Figure \ref{fig:P_start}) and almost does not increase as $\rho$ increases, whereas the mean distance $\mathbb{E}[d] = 1/\rho$ keeps reducing, thus increasing the chances for the AN to leave its serving IN's connectivity range and be misaligned. 
  \item $P_{\rm NL} $ increases when $V $ decreases since, when the VN is slower, it has less chances to leave the communication range of the IN  within the same time slot of duration $T_{\rm RTO}$.
\item $P_{\rm NL}$ increases when $T_{\rm RTO} $ decreases, since the AN  covers a shorter distance $ V \cdot T_{\rm RTO} $ moving at the same speed $ V$, thus reducing the chances to  leave its serving IN's communication~range.
\end{itemize}

\subsection{Mean Communication Duration}

The communication is "active" only  when the vehicle is both within the coverage range of the IN and correctly aligned, during the slot of duration $\T$. 
Constraining on the probability of having started the communication, the mean communication duration is: 
\begin{equation}
\mathbb{E}[T_{\rm comm}] =P_{\rm start}\cdot \Big[\mathbb{P}(T_{\rm L} > T_{\rm RTO})T_{\rm RTO}\:  + \:  \Big(1-\mathbb{P}(T_{\rm L} > T_{\rm RTO})\Big)\mathbb{E}[T_{\rm L}] \Big].
\end{equation}

\noindent In particular, if the  vehicle was already in the communication range at the beginning of the slot:
\begin{enumerate}
\item if the vehicle never overtakes its serving IN and never becomes misaligned within the slot (with probability $\mathbb{P}(T_{\rm L} > T_{\rm RTO})$), it communicates for the whole slot, so for a duration $\T$;
\item if, at some point, the vehicle  becomes misaligned with its IN, although being inside $\R$ during the slot (with probability $1-\mathbb{P}(T_{\rm L} > T_{\rm RTO})$), it communicates only for the time $\mathbb{E}[T_{\rm L}]<\T$ in which it was correctly able to be served (during this time window, the VN has covered a distance $d|d<\T V$, moving at speed $V$), where:
\begin{align}
&\mathbb{E}[T_{\rm L}] = \\\notag
&=\mathbb{E}[\text{Time in which the VN is aligned within } \R | \text{VN is connected but leaves } ] = \\ \notag
& = \dfrac{\mathbb{E}[{d|d<\T V | d < \R}]}{V}= \\ 
& = \frac{1}{V}\frac{1}{\mathbb{P}[d<\T V | d<\R]}\int_0^{\T V} \eta  p_d(\eta)d\eta \overset{(a)}   =  \\ \notag
&= \frac{1}{V}\frac{1-e^{-\rho\R}}{1-e^{-\rho\T V}}\frac{1-e^{-\rho\T V}(\rho\T V +1)}{\rho}\notag,
\end{align}
\end{enumerate}
where step $(a)$ is based on the fact that, in realistic vehicular environments, $\T V \ll \R$. 
In Figure \ref{fig:T}, we plot the \emph{communication duration ratio}, that is the ratio between the portion of time slot in which the VN is both within the communication range $\R$ of its serving IN and properly aligned ($\mathbb{E}[T_{\rm comm}]$) and the whole slot duration ($\T$) (i.e., $\mathbb{E}[T_{\rm comm}]/\T=1$ if the VN is connected for the whole time slot).
We observe that results agree with the considerations we made in the previous subsection and, in particular:

\begin{figure}[t!]
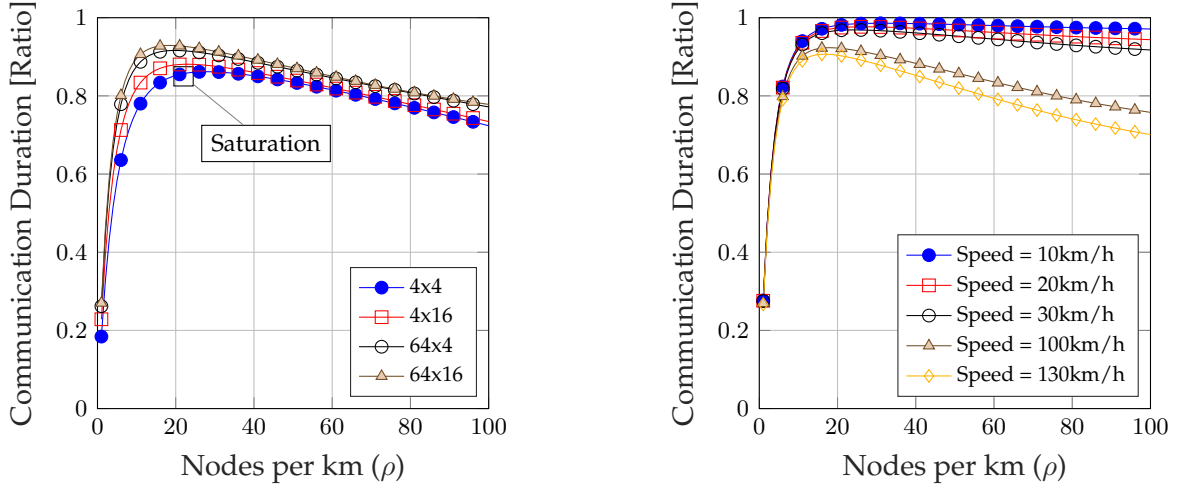
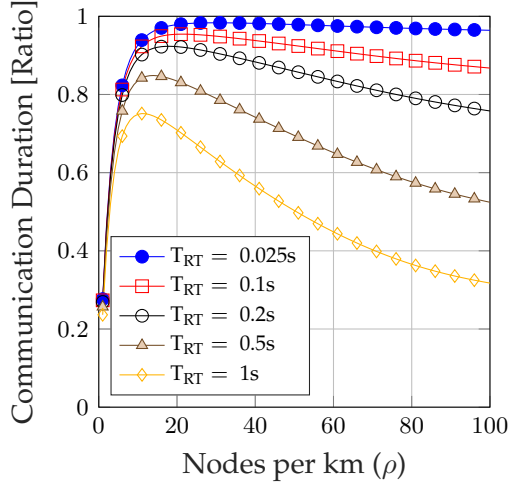

\centering
     \begin{subfigure}[t!]{0.45\textwidth}
     \centering
     		\setlength{\belowcaptionskip}{0cm}
      \iftikz
	\setlength{\belowcaptionskip}{0cm}
	\setlength\fwidth{0.7\columnwidth}
	\setlength\fheight{0.7\columnwidth}
	\input{./Figures/T_MIMO.tex}
	\else
   \includegraphics[width=\textwidth]{Figures/T_MIMO.eps}
    \fi
        \caption{$\mathbb{E}[T_{\rm L}]/\T$ with $V=90$ km/s and $\T = 200$ ms, for different MIMO configurations.}
   \end{subfigure} \hfill
             \begin{subfigure}[t!]{0.45\textwidth}
             		\setlength{\belowcaptionskip}{0cm}
             \iftikz
	\setlength{\belowcaptionskip}{0cm}
	\setlength\fwidth{0.7\columnwidth}
	\setlength\fheight{0.7\columnwidth}
	\input{./Figures/T_V.tex}
	\else
   \includegraphics[width=\textwidth]{Figures/T_V.eps}\fi
    \caption{$\mathbb{E}[T_{\rm L}]/\T$ with $\T = 200$ ms and MIMO configuration $64\times 16$, for different speeds.}
      \end{subfigure} \qquad
      \begin{subfigure}[t!]{0.45\textwidth}
      		\setlength{\belowcaptionskip}{0cm}
  \iftikz
	\setlength{\belowcaptionskip}{0cm}
	\setlength\fwidth{0.7\columnwidth}
	\setlength\fheight{0.7\columnwidth}
	\input{./Figures/T_T.tex}
	\else
   \includegraphics[width=\textwidth]{Figures/T_T.eps} \fi
    \caption{$\mathbb{E}[T_{\rm L}]/\T$ with $V=90$ km/h and MIMO configuration $64\times 16$, for different slot durations.}
      \end{subfigure}
\caption{Portion of time slot (of duration $\T$) in which the VN is both within the communication range $\R$ of its serving IN and properly aligned, when varying the nodes spatial density $\rho$.}
\label{fig:T}
\end{figure}

\begin{itemize}
\item $\mathbb{E}[T_{\rm comm}]/\T$ increases with $\rho$, for sparse networks (\textbf{$\rho$ small}), since the mean distance~$\mathbb{E}[d] = 1/\rho$ is relatively large and it is very unlikely for the AN to overtake its serving IN and become misaligned. 
The increasing behavior of the communication duration ratio is therefore caused by the increased probability of being within $\R$ at the beginning of the slot.
 \item $\mathbb{E}[T_{\rm comm}]/\T$ decreases with increasing values of$\rho $, for dense networks (\textbf{$\rho$ large}). In fact $P_{\rm start}$ has reached a quasi-steady state (see Figure \ref{fig:P_start}) and almost does not increase as $\rho$ increases, while the mean distance $\mathbb{E}[d] = 1/\rho$ keeps reducing, thus increasing the chances for the AN to overtake the IN and disconnect. 
  \item$\mathbb{E}[T_{\rm comm}]/\T$ increases  for lower values of $V $ and $\T$. Therefore, we assess that considering slower vehicular nodes or updating more frequently the AN-IN beam pair, respectively, reflects an increased average communication duration\footnote{Of course, more frequent beam-alignment updates has some overhead issues which must be~considered.}.
\end{itemize}

\section{Throughput Analysis}
\label{sec:analysis}

\begin{figure}[t!]
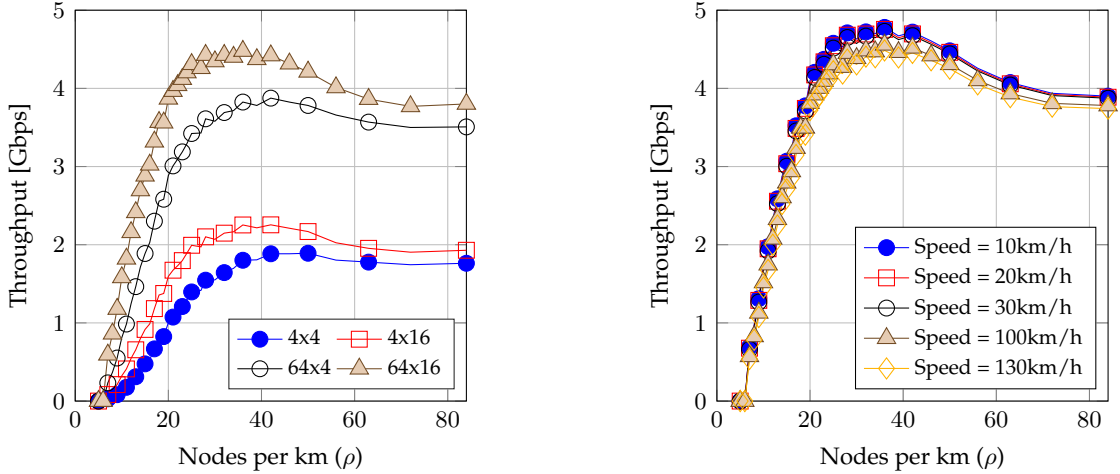
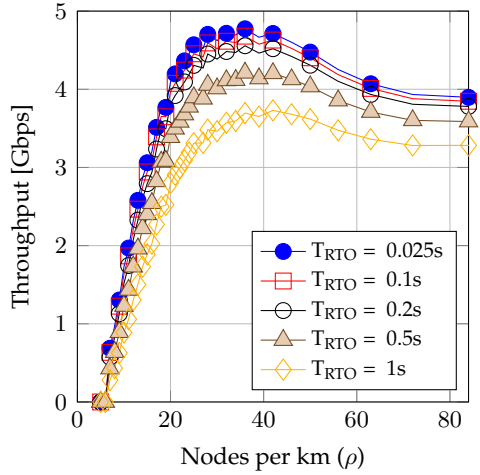

\centering
     \begin{subfigure}[t!]{0.45\textwidth}
     \centering
     		\setlength{\belowcaptionskip}{0cm}
      \iftikz
	\setlength{\belowcaptionskip}{0cm}
	\setlength\fwidth{0.7\columnwidth}
	\setlength\fheight{0.7\columnwidth}
	\input{./Figures/B_MIMO.tex}
	\else
   \includegraphics[width=\textwidth]{Figures/T_MIMO.eps}
    \fi
        \caption{Throughput with $V=90$ km/s and $\T = 200$ ms, for different MIMO configurations.}
   \end{subfigure} \hfill
             \begin{subfigure}[t!]{0.45\textwidth}
             		\setlength{\belowcaptionskip}{0cm}
             \iftikz
	\setlength{\belowcaptionskip}{0cm}
	\setlength\fwidth{0.7\columnwidth}
	\setlength\fheight{0.7\columnwidth}
	\input{./Figures/B_V.tex}
	\else
   \includegraphics[width=\textwidth]{Figures/T_V.eps}\fi
    \caption{Throughput  with $\T = 200$ ms and MIMO configuration $64\times 16$, for different speeds.}
      \end{subfigure} \qquad
      \begin{subfigure}[t!]{0.45\textwidth}
      		\setlength{\belowcaptionskip}{0cm}
  \iftikz
	\setlength{\belowcaptionskip}{0cm}
	\setlength\fwidth{0.7\columnwidth}
	\setlength\fheight{0.7\columnwidth}
	\input{./Figures/B_T.tex}
	\else
   \includegraphics[width=\textwidth]{Figures/T_T.eps} \fi
    \caption{Throughput  with $V=90$ km/h and MIMO configuration $64\times 16$, for different slot durations.}
      \end{subfigure}
\caption{Average throughput within a time slot of duration $\T$ when varying the nodes spatial density~$\rho$.}
\label{fig:B}
\end{figure}

A non-zero throughput can be perceived (within the slot of duration $\T$) only when the vehicle is within coverage and properly aligned with the infrastructure, that is for a time $\mathbb{E}[T_{\rm comm}]$. 
In this period, the vehicle perceives a rate $\mathbb{E}[R(d)]$ that depends on the mean distance $\mathbb{E}[d] = 1/\rho$, proportional to the nodes spatial density~$\rho$. The throughput is therefore defined as:

\begin{equation}
B = \mathbb{E}[R(d)] \cdot \frac{\mathbb{E}[T_{\rm comm}]}{T_{\rm RTO}}
\end{equation}

In Figure \ref{fig:B} and \ref{fig:hist}, we report the average throughput, when varying some system parameters. It is rather interesting to observe that, in all considered configurations, the throughput exhibits a similar pattern when varying the node density $\rho$. In particular:

\begin{figure}[t!]
\centering
             \iftikz
	\setlength{\belowcaptionskip}{0cm}
	\setlength\fwidth{0.89\columnwidth}
	\setlength\fheight{0.4\columnwidth}
	\input{./Figures/Hist_B.tex}
	\else
   \includegraphics[width=\textwidth]{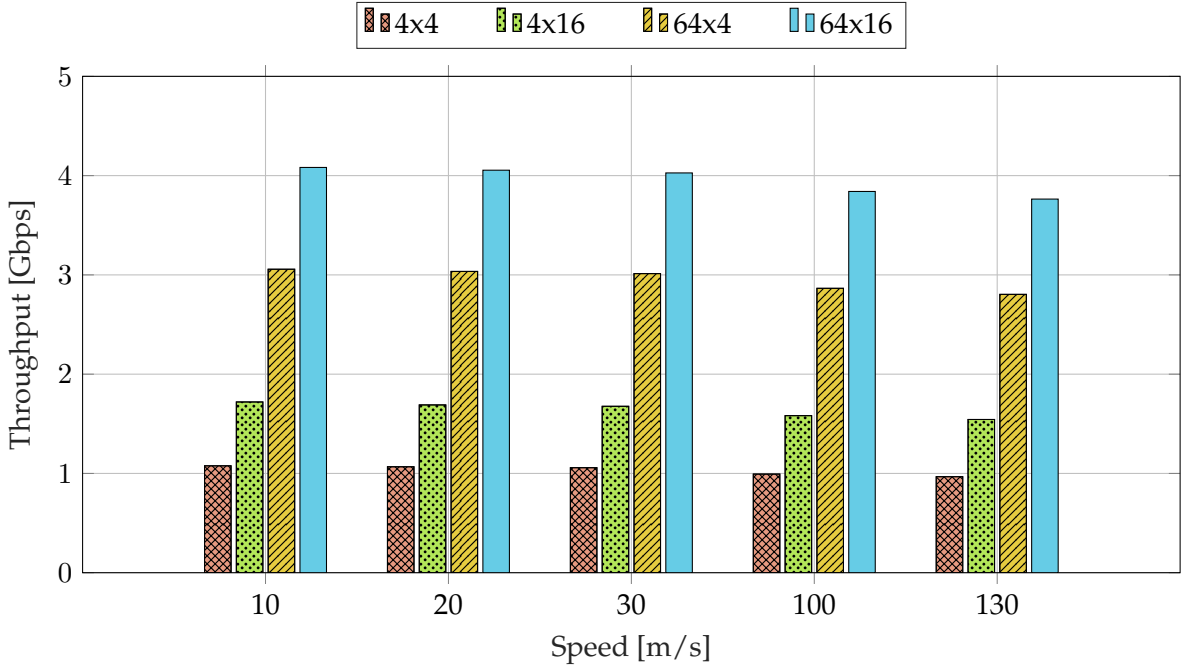}\fi 
\caption{Average throughput perceived by a vehicular node, considering a node spatial density $\rho=20$ nodes/km and $\T = 200$ ms, for different MIMO configurations and for different speed values $V$.}
\label{fig:hist}
\end{figure}

\begin{itemize}
\item We note that, as the node density is increased, the average distance between the AN and the IN decreases and, hence, the average bit rate experienced by the AN in case of connectivity becomes larger. On the other hand, the smaller coverage range will increase the probability of losing connectivity during a slot and will determine an increment of the frequency of handovers. 
\item For small values of $\rho$, $B$ initially increases with $\rho$. In this region, the SINR increases with $\rho$ because the reduction of the mean distance to the serving IN is more significant than the increase of the interference coming from the neighboring INs. 
Moreover, the distance between adjacent INs is still sufficiently large to allow for a loose beam alignment (thanks to the widening of the beam with the distance), so that the connectivity between the AN and the IN is maintained for a relatively large number of slots.  
\item After a certain value of $\rho$ (approximately 40 nodes/km in our scenario), $B$ starts decreasing. In this region, the interference from close-by INs becomes dominant and the perceived SINR degrades. 
Moreover, the closer the distance between the IN and the AN, the smaller the beam widening and, hence, the higher the risk of losing connectivity during a~slot.
\item the throughput grows as $V$ decreases since a slower AN is less likely to lose connectivity to the serving IN during a slot.
\item Similarly, the throughput grows as $\T$ decreases, because the beam alignment is repeated more frequently, reducing the disconnection time. However, the overhead (which is not accounted for in this simple analysis) would also increase, thus limiting or even nullifying the gain.  
\item Finally, as shown in Figure~\ref{fig:Rcomm_rho}, the throughput grows with the MIMO array size due to the increased achievable communication range $\R$ of the nodes. 
\end{itemize}

In conclusion, we have presented a preliminary connectivity and coverage study in a simple automotive scenario using mmWave communication link, and show how the performance of common directional beam tracking protocols can be improved by accounting for the specificities of the automotive scenario.

%\begin{figure}[t!]
%\centering
%     \begin{subfigure}[t!]{0.45\textwidth}
%     \centering
%     		\setlength{\belowcaptionskip}{0cm}
%      \iftikz
%	\setlength{\belowcaptionskip}{0cm}
%	\setlength\fwidth{0.8\columnwidth}
%	\setlength\fheight{0.8\columnwidth}
%	\input{./Figures/Hist_T.tex}
%	\else
%   \includegraphics[width=\textwidth]{Figures/hist1.eps}
%    \fi
%        \caption{Throughput with $V=90$ km/s and $\T = 200$ ms, for different MIMO configurations.}
%   \end{subfigure} \hfill
%             \begin{subfigure}[t!]{0.45\textwidth}
%             		\setlength{\belowcaptionskip}{0cm}
%             \iftikz
%	\setlength{\belowcaptionskip}{0cm}
%	\setlength\fwidth{0.8\columnwidth}
%	\setlength\fheight{0.8\columnwidth}
%	\input{./Figures/Hist_B.tex}
%	\else
%   \includegraphics[width=\textwidth]{Figures/hist2.eps}\fi
%    \caption{Throughput  with $\T = 200$ ms and MIMO configuration $64\times 16$, for different speeds.}
%      \end{subfigure}  
%\caption{Average throughput within a time slot of duration $\T$ when varying the nodes spatial density~$\rho$.}
%\label{fig:hist}
%\end{figure}

\bibliographystyle{IEEEtran}
\bibliography{biblio}

%\newpage
%
%\section{Main References}
%[1]: X. Zhang, F. Xie, W. Wang and M. Chatterjee, "TCP Throughput for Vehicle-to-Vehicle Communications,"  \textit{First International Conference on Communications and Networking in China}, Beijing, 2006.\\
%
%\noindent [2]: Rudack, M., M. Meincke, and M. Lott. "On the dynamics of ad hoc networks for inter vehicle communications (IVC)." in Proceedings of  \textit{ICWN}, 2002.\\
%
%\noindent [3]: M. Giordani, A. Zanella, and M. Zorzi, "Millimeter Wave Communication in Vehicular Networks: Challenges and Opportunities", accepted to \emph{International Conference on Modern Circuits and Systems Technologies (MOCAST)}, 2017.
\end{document}

%% file: Figures/Rcomm_rho.tex
\begin{tikzpicture}
\pgfplotsset{
tick label style={font=\footnotesize},
label style={font=\footnotesize},
legend style={font=\small}
}
\tikzset{every mark/.append style={scale=1.5}}

\begin{axis}[%
width=\fwidth,
height=\fheight,
at={(0\fwidth,0\fheight)},
scale only axis,
xmin=0,
xmax=90,
xlabel style={font=\color{white!15!black}},
xlabel={$\text{Nodes per km (}\rho\text{)}$},
ymin=20,
ymax=180,
ylabel style={font=\color{white!15!black}},
ylabel={$\text{R}_{\text{comm}}$},
axis background/.style={fill=white},
xmajorgrids,
ymajorgrids,
legend columns = 2,
legend style={legend cell align=left, align=left, draw=white!15!black},
cycle list={%
{blue,mark=*},
{red,mark=square},
{black,mark=o},
{brown!60!black,
mark options={fill=brown!40, scale = 2},
mark=triangle*}}
]
\addplot
  table[row sep=crcr]{%
1	107\\
11	79\\
21	63\\
31	53\\
41	45\\
51	37\\
61	33\\
71	29\\
81	25\\
91	23\\
};
\addlegendentry{4x4}

\addplot 
  table[row sep=crcr]{%
1	135\\
11	95\\
21	71\\
31	59\\
41	47\\
51	39\\
61	35\\
71	31\\
81	27\\
91	25\\
};
\addlegendentry{4x16}

\addplot 
  table[row sep=crcr]{%
1	157\\
11	119\\
21	93\\
31	75\\
41	63\\
51	53\\
61	47\\
71	43\\
81	39\\
91	35\\
};
\addlegendentry{64x4}

\addplot 
  table[row sep=crcr]{%
1	170\\
11	136\\
21	105\\
31	83\\
41	69\\
51	59\\
61	51\\
71	47\\
81	41\\
91	37\\
};
\addlegendentry{64x16}

\end{axis}
\end{tikzpicture}%

%% file: Figures/P_start.tex
\usetikzlibrary{spy}

\tikzstyle{every pin}=[fill=white,
draw=black,
font=\footnotesize]

\begin{tikzpicture}[spy using outlines=
	{circle, magnification=2, connect spies}]

\pgfplotsset{
tick label style={font=\footnotesize},
label style={font=\footnotesize},
legend style={font=\small}
}
\tikzset{every mark/.append style={scale=1.2}}
\begin{axis}[%
width=\fwidth,
height=\fheight,
at={(0\fwidth,0\fheight)},
scale only axis,
xmin=0,
xmax=100,
xlabel style={font=\color{white!15!black}},
xlabel={$\text{Nodes per km (}\rho\text{)}$},
xlabel style={font=\color{white!15!black}},
ymin=0,
ymax=1,
xmajorgrids,
ymajorgrids,
ylabel style={font=\color{white!15!black}},
ylabel={$\text{P}_{\text{start}}$},
axis background/.style={fill=white},
legend columns = 2,
legend pos =south west,
legend style={legend cell align=left, align=left, draw=white!15!black},
cycle list={%
{blue,mark=*,mark repeat = 5},
{red,mark=square,mark repeat = 5},
{black,mark=o,mark repeat = 5},
{brown!60!black,
mark options={fill=brown!40, scale = 2, mark repeat = 5},
mark=triangle*}}
]
\addplot
  table[row sep=crcr]{%
1	0.192381780193394\\
2	0.33871936457151\\
3	0.451656946452642\\
4	0.53999673457666\\
5	0.609963623708125\\
6	0.666025171998714\\
7	0.711432244877759\\
8	0.748581150917044\\
9	0.779259837723385\\
10	0.804817625196125\\
11	0.826283778281105\\
12	0.844451378469716\\
13	0.859937352347415\\
14	0.873225916452469\\
15	0.884700353204487\\
16	0.89466648544362\\
17	0.903370183382111\\
18	0.911010538947717\\
19	0.917749864777942\\
20	0.923721345055958\\
21	0.929034935001462\\
22	0.933781943468186\\
23	0.938038617610484\\
24	0.941868965703269\\
25	0.945326994219308\\
26	0.948458491506397\\
27	0.951302458230285\\
28	0.953892260914142\\
29	0.956256567123189\\
30	0.958420107485528\\
31	0.960404299640383\\
32	0.962227761519785\\
33	0.963906735486538\\
34	0.965455440321079\\
35	0.966886364541837\\
36	0.968210511812585\\
37	0.969437607052801\\
38	0.970576270185892\\
39	0.971634163131379\\
40	0.972618114592072\\
41	0.973534226345731\\
42	0.974387964076577\\
43	0.975184235239739\\
44	0.97592745601377\\
45	0.976621609041244\\
46	0.977270293368477\\
47	0.977876767759373\\
48	0.978443988364922\\
49	0.978974641570822\\
50	0.979471172714383\\
51	0.979935811253277\\
52	0.980370592878477\\
53	0.980777378988664\\
54	0.981157873880645\\
55	0.981513639957861\\
56	0.981846111214952\\
57	0.98215660521928\\
58	0.982446333778959\\
59	0.982716412460488\\
60	0.982967869096609\\
61	0.983201651405906\\
62	0.983418633829405\\
63	0.98361962367555\\
64	0.983805366653031\\
65	0.983976551860735\\
66	0.98413381629536\\
67	0.984277748929649\\
68	0.984408894407715\\
69	0.984527756398273\\
70	0.984634800641736\\
71	0.984730457722877\\
72	0.984815125597081\\
73	0.984889171895006\\
74	0.984952936027649\\
75	0.985006731111373\\
76	0.985050845730309\\
77	0.985085545551632\\
78	0.985111074807587\\
79	0.985127657656642\\
80	0.985135499434888\\
81	0.98513478780764\\
82	0.985125693830198\\
83	0.98510837292583\\
84	0.985082965788238\\
85	0.985049599215067\\
86	0.985008386878379\\
87	0.984959430037466\\
88	0.984902818198858\\
89	0.984838629727939\\
90	0.984766932416183\\
91	0.984687784007663\\
92	0.984601232688145\\
93	0.984507317539815\\
94	0.98440606896439\\
95	0.984297509077159\\
96	0.98418165207426\\
97	0.984058504575322\\
98	0.983928065943419\\
99	0.983790328584108\\
100	0.983645278225227\\
};
\addlegendentry{4x4}

\addplot
  table[row sep=crcr]{%
1	0.236387019124004\\
2	0.405169071195448\\
3	0.528079520960806\\
4	0.619231206550726\\
5	0.687979327023439\\
6	0.740645699806739\\
7	0.781579701326647\\
8	0.813824297007338\\
9	0.839542106609329\\
10	0.860292828246694\\
11	0.877216908956685\\
12	0.891159065039924\\
13	0.902752585884712\\
14	0.912477678580121\\
15	0.920702380770925\\
16	0.927711608185111\\
17	0.933728021434277\\
18	0.938927183500159\\
19	0.943448686509748\\
20	0.947404401579206\\
21	0.950884653799844\\
22	0.95396288597791\\
23	0.956699211252963\\
24	0.959143141440953\\
25	0.961335698661185\\
26	0.963311061773604\\
27	0.96509785918764\\
28	0.966720190846906\\
29	0.968198441327104\\
30	0.969549930720472\\
31	0.970789438728134\\
32	0.971929629025809\\
33	0.972981394719053\\
34	0.97395414099852\\
35	0.974856017539293\\
36	0.975694110468246\\
37	0.97647460163635\\
38	0.977202901322145\\
39	0.977883759242568\\
40	0.978521357771973\\
41	0.97911939050514\\
42	0.979681128696945\\
43	0.980209477633624\\
44	0.980707024610249\\
45	0.981176079884953\\
46	0.98161871173618\\
47	0.982036776552199\\
48	0.982431944722485\\
49	0.982805722970684\\
50	0.983159473662851\\
51	0.983494431537624\\
52	0.98381171823348\\
53	0.984112354929003\\
54	0.984397273363143\\
55	0.984667325461608\\
56	0.984923291761506\\
57	0.985165888797844\\
58	0.985395775591556\\
59	0.985613559358535\\
60	0.985819800542133\\
61	0.986015017257149\\
62	0.986199689221049\\
63	0.986374261237754\\
64	0.986539146290421\\
65	0.986694728292011\\
66	0.986841364535942\\
67	0.986979387883513\\
68	0.987109108719975\\
69	0.987230816706962\\
70	0.987344782355452\\
71	0.9874512584403\\
72	0.987550481274735\\
73	0.987642671860887\\
74	0.9877280369304\\
75	0.987806769887457\\
76	0.987879051665001\\
77	0.987945051503645\\
78	0.988004927661596\\
79	0.988058828062929\\
80	0.988106890890656\\
81	0.988149245130307\\
82	0.988186011069023\\
83	0.988217300754637\\
84	0.988243218418644\\
85	0.988263860866581\\
86	0.988279317838892\\
87	0.988289672345036\\
88	0.988295000973293\\
89	0.988295374178445\\
90	0.988290856549293\\
91	0.988281507057749\\
92	0.988267379291068\\
93	0.988248521668629\\
94	0.98822497764453\\
95	0.988196785897125\\
96	0.988163980506547\\
97	0.988126591121136\\
98	0.988084643113617\\
99	0.988038157727799\\
100	0.987987152216493\\
};
\addlegendentry{4x16}

\addplot 
  table[row sep=crcr]{%
1	0.269550051889437\\
2	0.456713139761037\\
3	0.588980126146159\\
4	0.684018017138231\\
5	0.753383660062461\\
6	0.804765064314041\\
7	0.843358669941127\\
8	0.87273027371261\\
9	0.895362023361111\\
10	0.913005378601381\\
11	0.926912250739859\\
12	0.937988506859445\\
13	0.94689728777471\\
14	0.954129439802429\\
15	0.960052115305156\\
16	0.964942700045025\\
17	0.969012760955202\\
18	0.9724251293271\\
19	0.975306210739911\\
20	0.977754941396554\\
21	0.979849364838728\\
22	0.981651504056466\\
23	0.983211001386161\\
24	0.984567859889421\\
25	0.985754524047648\\
26	0.986797470748933\\
27	0.987718434498787\\
28	0.988535357400941\\
29	0.989263130568766\\
30	0.989914176404112\\
31	0.99049890866623\\
32	0.991026098094854\\
33	0.991503164601436\\
34	0.991936412033967\\
35	0.99233121778004\\
36	0.992692186661449\\
37	0.993023276447737\\
38	0.993327900699249\\
39	0.993609013413542\\
40	0.993869178997784\\
41	0.994110630354456\\
42	0.994335317296144\\
43	0.994544947058948\\
44	0.99474101833383\\
45	0.994924849959165\\
46	0.995097605199178\\
47	0.995260312359122\\
48	0.995413882349239\\
49	0.99555912369825\\
50	0.995696755427556\\
51	0.995827418124946\\
52	0.995951683497918\\
53	0.996070062638992\\
54	0.99618301319638\\
55	0.996290945611457\\
56	0.996394228558219\\
57	0.9964931936983\\
58	0.9965881398472\\
59	0.99667933663259\\
60	0.996767027713182\\
61	0.996851433616411\\
62	0.996932754244542\\
63	0.997011171091634\\
64	0.997086849207734\\
65	0.997159938941545\\
66	0.997230577488533\\
67	0.99729889026776\\
68	0.997364992147642\\
69	0.997428988538175\\
70	0.99749097636494\\
71	0.997551044938242\\
72	0.997609276729098\\
73	0.997665748062359\\
74	0.997720529736019\\
75	0.997773687574705\\
76	0.99782528292441\\
77	0.997875373094736\\
78	0.997924011754205\\
79	0.997971249283585\\
80	0.998017133091651\\
81	0.998061707897302\\
82	0.998105015981557\\
83	0.998147097412588\\
84	0.998187990246592\\
85	0.998227730707054\\
86	0.998266353344643\\
87	0.99830389117981\\
88	0.998340375829892\\
89	0.998375837622389\\
90	0.998410305695879\\
91	0.998443808089909\\
92	0.998476371825054\\
93	0.998508022974218\\
94	0.998538786726149\\
95	0.998568687442016\\
96	0.998597748705854\\
97	0.998625993369544\\
98	0.998653443592978\\
99	0.998680120879948\\
100	0.99870604611028\\
};
\addlegendentry{64x4}

\addplot
  table[row sep=crcr]{%
1	0.277588036954274\\
2	0.470894180225896\\
3	0.607287680571618\\
4	0.70475121516606\\
5	0.775251954140798\\
6	0.826852680674321\\
7	0.865050720437025\\
8	0.893637592109494\\
9	0.915257518425388\\
10	0.931774470845802\\
11	0.944516074505123\\
12	0.954437432881694\\
13	0.962232311896239\\
14	0.968409358941148\\
15	0.973344862185506\\
16	0.97731961537986\\
17	0.980544911606274\\
18	0.983181033471861\\
19	0.985350518047173\\
20	0.987147751761937\\
21	0.98864596609619\\
22	0.989902377596059\\
23	0.990961992690711\\
24	0.991860444529589\\
25	0.99262612291567\\
26	0.993281784327223\\
27	0.993845776924669\\
28	0.994332978537384\\
29	0.994755519300844\\
30	0.995123341706334\\
31	0.995444637153456\\
32	0.995726188145644\\
33	0.995973637982004\\
34	0.996191704429546\\
35	0.996384349880105\\
36	0.996554917529323\\
37	0.996706240890768\\
38	0.996840732281636\\
39	0.996960454645915\\
40	0.997067180113063\\
41	0.997162437949345\\
42	0.99724755398893\\
43	0.997323683191217\\
44	0.997391836628631\\
45	0.997452903942121\\
46	0.997507672092354\\
47	0.997556841069864\\
48	0.997601037097235\\
49	0.997640823753055\\
50	0.997676711364986\\
51	0.997709164953437\\
52	0.997738610954406\\
53	0.997765442907456\\
54	0.997790026260358\\
55	0.997812702414098\\
56	0.997833792109328\\
57	0.997853598237056\\
58	0.997872408141564\\
59	0.99789049547156\\
60	0.997908121626064\\
61	0.997925536833919\\
62	0.997942980899974\\
63	0.997960683646556\\
64	0.997978865075662\\
65	0.997997735275202\\
66	0.998017494091479\\
67	0.998038330589755\\
68	0.998060422325127\\
69	0.99808393444694\\
70	0.998109018661452\\
71	0.998135812079367\\
72	0.99816443597705\\
73	0.998194994502586\\
74	0.998227573360283\\
75	0.998262238509503\\
76	0.998299034915862\\
77	0.998337985394495\\
78	0.998379089586311\\
79	0.998422323108662\\
80	0.998467636921504\\
81	0.998514956948859\\
82	0.998564183992977\\
83	0.998615193975031\\
84	0.998667838531357\\
85	0.998721945988109\\
86	0.998777322729841\\
87	0.998833754968942\\
88	0.998891010913185\\
89	0.998948843318145\\
90	0.999006992400033\\
91	0.999065189073028\\
92	0.999123158463696\\
93	0.999180623644084\\
94	0.999237309514895\\
95	0.999292946761388\\
96	0.999347275797597\\
97	0.999400050609674\\
98	0.999451042406983\\
99	0.999500042990285\\
100	0.999546867750179\\
};
\addlegendentry{64x16}

 \coordinate (spypoint) at (axis cs:40,1);
  \coordinate (magnifyglass) at (axis cs:77,0.5);

%\node[coordinate,pin=below right :{Reached saturation point}]
%at (axis cs:20,0.923721345055958) {};

\end{axis}

\spy [blue, size=2.5cm] on (spypoint)
   in node[fill=white] at (magnifyglass);
\end{tikzpicture}%

%% file: Figures/P_NL_MIMO.tex

\tikzstyle{every pin}=[fill=white,
draw=black,
font=\footnotesize]
\begin{tikzpicture}
\pgfplotsset{
tick label style={font=\footnotesize},
label style={font=\footnotesize},
legend style={font=\scriptsize}
}
\tikzset{every mark/.append style={scale=1.2}}
\begin{axis}[%
width=\fwidth,
height=\fheight,
at={(0\fwidth,0\fheight)},
scale only axis,
xmin=0,
xmax=100,
xlabel style={font=\color{white!15!black}},
xlabel={$\text{Nodes per km (}\rho\text{)}$},
xlabel style={font=\color{white!15!black}},
ymin=0,
ymax=1,
xmajorgrids,
ymajorgrids,
ylabel style={font=\color{white!15!black}},
ylabel={$\text{P}_{\text{NL}}$},
axis background/.style={fill=white},
legend columns = 1,
legend pos =south east,
legend style={legend cell align=left, align=left, draw=white!15!black},
cycle list={%
{blue,mark=*, mark repeat = 5},
{red,mark=square, mark repeat = 5},
{black,mark=o, mark repeat = 5},
{brown!60!black,
mark options={fill=brown!40, scale = 1.2, mark repeat = 5},
mark=triangle*}}
]
\addplot
  table[row sep=crcr]{%
1	0.186940107737215\\
2	0.327257658618421\\
3	0.43390801364802\\
4	0.515874121980608\\
5	0.579483894117113\\
6	0.62926146312354\\
7	0.668486573497429\\
8	0.699568028570305\\
9	0.72429674256672\\
10	0.744019634725436\\
11	0.759760720165586\\
12	0.77230649280116\\
13	0.782266846476318\\
14	0.790119031912072\\
15	0.796239711552494\\
16	0.80092857131198\\
17	0.804425879925074\\
18	0.806925666262641\\
19	0.808585693797022\\
20	0.809535072881411\\
21	0.809880115819548\\
22	0.809708874004132\\
23	0.809094678825326\\
24	0.808098923876301\\
25	0.80677326520809\\
26	0.805161372148375\\
27	0.803300328746802\\
28	0.801221761924948\\
29	0.798952754555442\\
30	0.796516588313142\\
31	0.793933351045211\\
32	0.791220435741189\\
33	0.788392952328151\\
34	0.78546406901614\\
35	0.782445296441669\\
36	0.779346725155189\\
37	0.776177224887975\\
38	0.772944612377133\\
39	0.769655793220461\\
40	0.766316882196926\\
41	0.762933305663706\\
42	0.759509888981135\\
43	0.756050931387068\\
44	0.752560270315009\\
45	0.749041336804411\\
46	0.745497203370472\\
47	0.741930625471369\\
48	0.738344077523165\\
49	0.734739784258322\\
50	0.731119748096628\\
51	0.727485773092138\\
52	0.723839485932516\\
53	0.720182354394507\\
54	0.716515703598674\\
55	0.712840730355776\\
56	0.709158515854574\\
57	0.705470036904969\\
58	0.701776175920124\\
59	0.698077729795574\\
60	0.694375417821648\\
61	0.690669888746965\\
62	0.686961727095084\\
63	0.68325145882287\\
64	0.679539556397625\\
65	0.675826443360146\\
66	0.672112498432324\\
67	0.668398059220601\\
68	0.664683425560228\\
69	0.660968862539797\\
70	0.657254603240777\\
71	0.653540851222608\\
72	0.649827782780365\\
73	0.64611554899883\\
74	0.642404277624078\\
75	0.638694074771307\\
76	0.634985026485512\\
77	0.631277200169777\\
78	0.627570645894346\\
79	0.623865397598183\\
80	0.620161474193495\\
81	0.616458880582571\\
82	0.612757608595307\\
83	0.60905763785492\\
84	0.605358936578571\\
85	0.601661462318964\\
86	0.597965162652323\\
87	0.594269975817664\\
88	0.590575831311759\\
89	0.586882650443768\\
90	0.583190346853125\\
91	0.579498826993937\\
92	0.575807990588807\\
93	0.572117731054764\\
94	0.568427935903685\\
95	0.564738487119413\\
96	0.561049261513543\\
97	0.557360131061669\\
98	0.55367096322176\\
99	0.549981621236109\\
100	0.546291964418261\\
};
\addlegendentry{4x4}

\addplot
  table[row sep=crcr]{%
1	0.230849691291864\\
2	0.393463477690844\\
3	0.509989869430988\\
4	0.594750908226446\\
5	0.657201467421532\\
6	0.703705155565119\\
7	0.738623126774417\\
8	0.764995184717412\\
9	0.784973964960308\\
10	0.800106667823897\\
11	0.811520947883556\\
12	0.820049509662952\\
13	0.826314867720807\\
14	0.83078781992255\\
15	0.833828323055386\\
16	0.835714426206277\\
17	0.836662994631685\\
18	0.836844720722985\\
19	0.836395113061975\\
20	0.835422622692932\\
21	0.834014710248048\\
22	0.83224241716129\\
23	0.830163839807842\\
24	0.827826791780459\\
25	0.82527086017829\\
26	0.822529005850321\\
27	0.81962881773407\\
28	0.816593502856145\\
29	0.813442672872882\\
30	0.810192972928463\\
31	0.806858587499606\\
32	0.803451649663509\\
33	0.79998257408103\\
34	0.796460329369155\\
35	0.792892662043475\\
36	0.78928628155202\\
37	0.785647013885088\\
38	0.781979929676646\\
39	0.778289451497403\\
40	0.774579444092893\\
41	0.770853290578654\\
42	0.767113957021274\\
43	0.763364047372828\\
44	0.759605850359739\\
45	0.755841379634628\\
46	0.752072408265178\\
47	0.748300498445259\\
48	0.74452702716084\\
49	0.740753208419267\\
50	0.736980112549423\\
51	0.733208682997544\\
52	0.729439750975549\\
53	0.725674048262647\\
54	0.721912218414635\\
55	0.718154826596725\\
56	0.714402368223629\\
57	0.71065527656375\\
58	0.706913929441756\\
59	0.703178655154819\\
60	0.699449737701761\\
61	0.695727421410737\\
62	0.692011915039522\\
63	0.688303395412627\\
64	0.684602010651044\\
65	0.680907883043191\\
66	0.677221111599428\\
67	0.673541774327193\\
68	0.66986993025916\\
69	0.66620562126284\\
70	0.662548873656614\\
71	0.658899699654115\\
72	0.65525809865632\\
73	0.651624058408388\\
74	0.64799755603629\\
75	0.644378558976542\\
76	0.640767025810799\\
77	0.637162907015738\\
78	0.633566145637472\\
79	0.629976677898681\\
80	0.626394433745756\\
81	0.622819337342405\\
82	0.619251307515491\\
83	0.615690258158215\\
84	0.612136098595194\\
85	0.60858873391352\\
86	0.60504806526339\\
87	0.601513990131568\\
88	0.597986402590541\\
89	0.594465193525962\\
90	0.590950250844655\\
91	0.58744145966527\\
92	0.583938702493398\\
93	0.580441859382807\\
94	0.576950808084262\\
95	0.573465424183248\\
96	0.569985581227772\\
97	0.566511150847311\\
98	0.563042002863827\\
99	0.559578005395726\\
100	0.556119024955502\\
};
\addlegendentry{4x16}

\addplot 
  table[row sep=crcr]{%
1	0.263943012678418\\
2	0.44483469999788\\
3	0.570666495114191\\
4	0.659342776758819\\
5	0.722521334707859\\
6	0.76792190372337\\
7	0.800739031974074\\
8	0.824521708583521\\
9	0.841729995800672\\
10	0.854092772455661\\
11	0.862841074445327\\
12	0.868862450530177\\
13	0.872804449596566\\
14	0.875144897904885\\
15	0.876240211020058\\
16	0.87635899602314\\
17	0.875705685139956\\
18	0.874437336527983\\
19	0.872675700313003\\
20	0.870515969340668\\
21	0.868033185246967\\
22	0.865286970330814\\
23	0.862325052946266\\
24	0.859185915766107\\
25	0.855900800938601\\
26	0.85249523987118\\
27	0.848990228870405\\
28	0.845403138962627\\
29	0.841748424745236\\
30	0.838038180238385\\
31	0.834282577476036\\
32	0.830490214647356\\
33	0.826668394036235\\
34	0.822823345148695\\
35	0.818960404798205\\
36	0.815084163204509\\
37	0.811198583113525\\
38	0.807307097391216\\
39	0.803412689357509\\
40	0.799517959215108\\
41	0.795625179224717\\
42	0.791736339732395\\
43	0.787853187729141\\
44	0.78397725928918\\
45	0.780109906970731\\
46	0.776252323055208\\
47	0.772405559335648\\
48	0.768570544033368\\
49	0.764748096316213\\
50	0.760938938806782\\
51	0.757143708400361\\
52	0.753362965656649\\
53	0.749597202984074\\
54	0.745846851798508\\
55	0.74211228880795\\
56	0.738393841549811\\
57	0.734691793286978\\
58	0.731006387351844\\
59	0.727337831013476\\
60	0.723686298931404\\
61	0.720051936249816\\
62	0.716434861377793\\
63	0.712835168494485\\
64	0.709252929812379\\
65	0.705688197627081\\
66	0.702141006177961\\
67	0.698611373340672\\
68	0.695099302169628\\
69	0.691604782306147\\
70	0.688127791265892\\
71	0.684668295617474\\
72	0.681226252062641\\
73	0.677801608427165\\
74	0.674394304570478\\
75	0.671004273221183\\
76	0.667631440744754\\
77	0.664275727849077\\
78	0.660937050232873\\
79	0.657615319181545\\
80	0.654310442114548\\
81	0.651022323087984\\
82	0.647750863255809\\
83	0.64449596129271\\
84	0.641257513781495\\
85	0.638035415567553\\
86	0.634829560082788\\
87	0.631639839641222\\
88	0.628466145708287\\
89	0.625308369145721\\
90	0.622166400433795\\
91	0.61904012987252\\
92	0.615929447763347\\
93	0.612834244572792\\
94	0.609754411079296\\
95	0.606689838504566\\
96	0.60364041863057\\
97	0.600606043903245\\
98	0.597586607523954\\
99	0.594582003529637\\
100	0.591592126862536\\
};
\addlegendentry{64x4}

\addplot 
  table[row sep=crcr]{%
1	0.271964426451716\\
2	0.458970987704207\\
3	0.588916479369557\\
4	0.68003679880875\\
5	0.744405937253754\\
6	0.79011758938409\\
7	0.822662658176994\\
8	0.845809722672564\\
9	0.862174502542762\\
10	0.87359241102338\\
11	0.881364999151208\\
12	0.88642473643028\\
13	0.889446338785607\\
14	0.890922743768238\\
15	0.89121746748454\\
16	0.890601028082795\\
17	0.889276518045464\\
18	0.887397718401615\\
19	0.885082041183389\\
20	0.882419854494097\\
21	0.879481256123879\\
22	0.876321032868094\\
23	0.872982319509902\\
24	0.869499318663293\\
25	0.865899337275388\\
26	0.86220432230418\\
27	0.858432026749535\\
28	0.854596900985935\\
29	0.85071077859584\\
30	0.846783407474291\\
31	0.842822863697645\\
32	0.838835876019708\\
33	0.834828081829978\\
34	0.830804230246669\\
35	0.82676834420295\\
36	0.82272385054974\\
37	0.81867368507898\\
38	0.814620377777911\\
39	0.81056612242033\\
40	0.806512833685313\\
41	0.802462194294361\\
42	0.798415694120706\\
43	0.79437466280972\\
44	0.790340297127585\\
45	0.786313684004461\\
46	0.782295820041742\\
47	0.778287628098165\\
48	0.774289971447025\\
49	0.770303665899337\\
50	0.76632949020999\\
51	0.762368195021508\\
52	0.758420510549712\\
53	0.754487153174897\\
54	0.750568831069065\\
55	0.746666248962872\\
56	0.742780112134018\\
57	0.738911129680924\\
58	0.735060017131016\\
59	0.731227498421182\\
60	0.727414307278557\\
61	0.723621188022414\\
62	0.719848895802297\\
63	0.716098196283433\\
64	0.712369864787759\\
65	0.708664684897448\\
66	0.704983446527529\\
67	0.701326943474974\\
68	0.69769597045343\\
69	0.69409131962547\\
70	0.690513776647819\\
71	0.686964116249423\\
72	0.683443097367287\\
73	0.679951457870761\\
74	0.676489908911245\\
75	0.673059128940931\\
76	0.669659757451263\\
77	0.666292388488923\\
78	0.662957564014369\\
79	0.659655767174966\\
80	0.656387415571504\\
81	0.653152854603061\\
82	0.649952350980738\\
83	0.646786086505395\\
84	0.643654152208086\\
85	0.640556542954224\\
86	0.637493152613369\\
87	0.634463769895907\\
88	0.631468074955488\\
89	0.628505636851992\\
90	0.625575911963732\\
91	0.622678243429782\\
92	0.61981186169351\\
93	0.616975886206802\\
94	0.614169328341161\\
95	0.611391095536834\\
96	0.608639996704758\\
97	0.605914748878367\\
98	0.603213985093604\\
99	0.60053626345593\\
100	0.5978800773331\\
};
\addlegendentry{64x16}

%\node[coordinate,pin={[pin anchor=south west,pin edge pin anchor=south west,pin distance=6ex]90:Z-Value}]
%at (axis cs:20,0.8) {};
\node[pin=290:Saturation,draw=black]
at (axis cs:20,0.8) {};

\end{axis}
\end{tikzpicture}%

%% file: Figures/P_NL_V.tex
\definecolor{mycolor1}{rgb}{0.00000,0.75000,0.75000}%
\begin{tikzpicture}
\pgfplotsset{
tick label style={font=\footnotesize},
label style={font=\footnotesize},
legend style={font=\scriptsize}
}
\tikzset{every mark/.append style={scale=1.2}}
\begin{axis}[%
width=\fwidth,
height=\fheight,
at={(0\fwidth,0\fheight)},
scale only axis,
xmin=0,
xmax=100,
xlabel style={font=\color{white!15!black}},
xlabel={$\text{Nodes per km (}\rho\text{)}$},
xlabel style={font=\color{white!15!black}},
ymin=0,
ymax=1,
xmajorgrids,
ymajorgrids,
ylabel style={font=\color{white!15!black}},
ylabel={$\text{P}_{\text{NL}}$},
axis background/.style={fill=white},
legend columns = 1,
legend pos =south east,
legend style={legend cell align=left, align=left, draw=white!15!black},
cycle list={%
{blue,mark=*, mark repeat = 5},
{red,mark=square, mark repeat = 5},
{black,mark=o, mark repeat = 5},
{brown!60!black,
mark options={fill=brown!40, scale = 1.2, mark repeat = 5},
mark=triangle*}}
]
\addplot 
  table[row sep=crcr]{%
1	0.276961801897378\\
2	0.469563485432787\\
3	0.605232801261175\\
4	0.701980696401566\\
5	0.771786414214077\\
6	0.822716370271564\\
7	0.860267340773037\\
8	0.888228420982106\\
9	0.909240757174916\\
10	0.925165210005541\\
11	0.937326548381167\\
12	0.946677386230339\\
13	0.953909387283476\\
14	0.959529460589815\\
15	0.963912477749496\\
16	0.967338090570943\\
17	0.970016679310817\\
18	0.972107801988331\\
19	0.97373342408242\\
20	0.974987483818461\\
21	0.975942863403383\\
22	0.976656509078116\\
23	0.977173219774546\\
24	0.977528470951527\\
25	0.977750534116715\\
26	0.97786207854139\\
27	0.977881389659958\\
28	0.977823301816029\\
29	0.977699916754538\\
30	0.977521160403157\\
31	0.977295216857141\\
32	0.977028868566723\\
33	0.976727764467527\\
34	0.976396632447991\\
35	0.976039448586237\\
36	0.975659572636411\\
37	0.975259857031865\\
38	0.974842735005083\\
39	0.974410292161067\\
40	0.973964324878839\\
41	0.973506388179459\\
42	0.973037835132533\\
43	0.972559849435462\\
44	0.972073472459716\\
45	0.971579625793224\\
46	0.97107913010009\\
47	0.970572720955273\\
48	0.970061062182474\\
49	0.969544757120817\\
50	0.969024358163994\\
51	0.968500374850068\\
52	0.967973280727454\\
53	0.967443519180207\\
54	0.966911508361405\\
55	0.96637764535569\\
56	0.965842309669429\\
57	0.965305866128698\\
58	0.964768667250458\\
59	0.964231055140329\\
60	0.963693362960805\\
61	0.963155916006141\\
62	0.962619032414308\\
63	0.962083023541936\\
64	0.961548194025042\\
65	0.961014841546284\\
66	0.960483256328372\\
67	0.959953720373083\\
68	0.959426506465817\\
69	0.958901876966852\\
70	0.958380082412153\\
71	0.957861359948829\\
72	0.957345931632862\\
73	0.956834002619506\\
74	0.956325759279718\\
75	0.955821367278809\\
76	0.955320969656339\\
77	0.954824684948657\\
78	0.954332605397565\\
79	0.953844795289959\\
80	0.953361289473982\\
81	0.952882092097002\\
82	0.952407175609474\\
83	0.951936480076391\\
84	0.951469912834414\\
85	0.951007348527973\\
86	0.950548629551486\\
87	0.950093566917536\\
88	0.949641941562373\\
89	0.949193506090645\\
90	0.948747986951052\\
91	0.948305087023835\\
92	0.947864488590036\\
93	0.947425856641656\\
94	0.946988842481485\\
95	0.946553087552023\\
96	0.946118227424855\\
97	0.945683895875504\\
98	0.945249728964545\\
99	0.94481536904386\\
100	0.944380468607616\\
};
\addlegendentry{Speed = 10km/h}

\addplot
  table[row sep=crcr]{%
1	0.276335914652224\\
2	0.468234268368336\\
3	0.60318134389717\\
4	0.699216327509715\\
5	0.768330487429386\\
6	0.818593824616158\\
7	0.855502527017299\\
8	0.882843237270431\\
9	0.903254004646374\\
10	0.918592565474911\\
11	0.930180824274377\\
12	0.938968901160419\\
13	0.945646356139495\\
14	0.950718360220021\\
15	0.954558369911639\\
16	0.957444897260948\\
17	0.959587412248927\\
18	0.961144750998996\\
19	0.962238310082679\\
20	0.962961582101842\\
21	0.96338710241084\\
22	0.963571549175225\\
23	0.963559515856181\\
24	0.96338632204449\\
25	0.963080122577428\\
26	0.962663500957411\\
27	0.962154681157044\\
28	0.96156845513814\\
29	0.960916897212266\\
30	0.96020991756649\\
31	0.959455693692031\\
32	0.958661008573587\\
33	0.957831517266886\\
34	0.956971958168337\\
35	0.956086321337352\\
36	0.955177983294091\\
37	0.954249815514318\\
38	0.953304272184829\\
39	0.952343461527034\\
40	0.951369204040067\\
41	0.950383080283113\\
42	0.949386470253881\\
43	0.948380585985318\\
44	0.947366498644966\\
45	0.94634516115794\\
46	0.945317427168071\\
47	0.944284066989243\\
48	0.943245781070429\\
49	0.942203211395912\\
50	0.941156951160791\\
51	0.940107552996717\\
52	0.939055535970455\\
53	0.938001391535617\\
54	0.936945588583769\\
55	0.935888577713402\\
56	0.934830794812766\\
57	0.933772664034288\\
58	0.932714600223501\\
59	0.931657010853407\\
60	0.930600297505659\\
61	0.929544856932308\\
62	0.928491081726026\\
63	0.927439360622267\\
64	0.926390078453734\\
65	0.925343615775496\\
66	0.924300348178104\\
67	0.923260645305955\\
68	0.922224869598834\\
69	0.921193374775908\\
70	0.920166504083488\\
71	0.919144588330301\\
72	0.918127943736973\\
73	0.917116869629584\\
74	0.916111646010586\\
75	0.915112531043801\\
76	0.914119758493601\\
77	0.913133535161562\\
78	0.912154038366616\\
79	0.911181413517079\\
80	0.910215771824462\\
81	0.909257188209796\\
82	0.908305699453028\\
83	0.907361302634754\\
84	0.906423953917191\\
85	0.905493567707556\\
86	0.904570016242167\\
87	0.903653129623307\\
88	0.90274269633356\\
89	0.901838464243833\\
90	0.90094014212184\\
91	0.900047401637751\\
92	0.899159879853083\\
93	0.898277182168211\\
94	0.897398885693311\\
95	0.896524542997566\\
96	0.895653686182337\\
97	0.894785831216189\\
98	0.893920482463362\\
99	0.89305713733289\\
100	0.892195290973218\\
};
\addlegendentry{Speed = 20km/h}

\addplot
  table[row sep=crcr]{%
1	0.275710375025637\\
2	0.466906527391534\\
3	0.601133302781107\\
4	0.696458094839297\\
5	0.764884147120604\\
6	0.814484997901999\\
7	0.850756207109267\\
8	0.877481934600294\\
9	0.897297111170637\\
10	0.91205633439399\\
11	0.923078635322002\\
12	0.931311635071329\\
13	0.937442787458639\\
14	0.941975524811897\\
15	0.945281889077353\\
16	0.947639253760319\\
17	0.94925618014882\\
18	0.950290784189621\\
19	0.950863895253293\\
20	0.951068561917581\\
21	0.950976974120599\\
22	0.950645543196204\\
23	0.950118658174085\\
24	0.949431483611426\\
25	0.948612058311823\\
26	0.947682880458903\\
27	0.946662112840166\\
28	0.945564505152863\\
29	0.944402104246637\\
30	0.943184804406453\\
31	0.94192077623667\\
32	0.940616802862526\\
33	0.939278544963542\\
34	0.937910750852585\\
35	0.936517423889313\\
36	0.935101956593535\\
37	0.933667238634545\\
38	0.932215744223517\\
39	0.930749603187507\\
40	0.929270659053194\\
41	0.927780516741424\\
42	0.926280581914524\\
43	0.924772093586375\\
44	0.923256151269829\\
45	0.921733737674418\\
46	0.92020573776228\\
47	0.918672954808788\\
48	0.917136123986732\\
49	0.915595923891499\\
50	0.914052986343851\\
51	0.912507904742094\\
52	0.910961241183368\\
53	0.909413532531757\\
54	0.907865295576888\\
55	0.906317031399082\\
56	0.904769229034694\\
57	0.90322236851703\\
58	0.901676923353431\\
59	0.900133362487161\\
60	0.898592151783138\\
61	0.89705375506898\\
62	0.895518634756954\\
63	0.893987252068033\\
64	0.892460066876184\\
65	0.890937537189073\\
66	0.889420118280439\\
67	0.887908261489449\\
68	0.886402412703117\\
69	0.884903010539497\\
70	0.883410484251552\\
71	0.881925251374389\\
72	0.880447715141799\\
73	0.878978261701651\\
74	0.877517257163527\\
75	0.876065044515998\\
76	0.874621940454891\\
77	0.873188232167745\\
78	0.871764174123176\\
79	0.870349984916975\\
80	0.868945844229218\\
81	0.867551889948416\\
82	0.866168215519541\\
83	0.864794867572556\\
84	0.863431843886735\\
85	0.862079091743473\\
86	0.860736506716418\\
87	0.859403931942644\\
88	0.858081157912102\\
89	0.856767922805037\\
90	0.855463913398329\\
91	0.85416876655216\\
92	0.852882071278137\\
93	0.851603371379277\\
94	0.85033216864149\\
95	0.849067926545453\\
96	0.847810074457693\\
97	0.84655801225029\\
98	0.845311115290422\\
99	0.844068739734122\\
100	0.842830228053378\\
};
\addlegendentry{Speed = 30km/h}

\addplot 
  table[row sep=crcr]{%
1	0.271341314755197\\
2	0.457653533545519\\
3	0.586892193226397\\
4	0.677321139991422\\
5	0.741025961812693\\
6	0.786103525428365\\
7	0.818043800865431\\
8	0.840612648178803\\
9	0.856422493938521\\
10	0.867305487637924\\
11	0.874560219454061\\
12	0.879116599715075\\
13	0.881647196930844\\
14	0.882643181424063\\
15	0.882466634263486\\
16	0.881386919369819\\
17	0.879606207349055\\
18	0.87727754682709\\
19	0.874517770354668\\
20	0.871416789045497\\
21	0.868044341256613\\
22	0.864454931698072\\
23	0.860691474166413\\
24	0.85678799839842\\
25	0.852771676236188\\
26	0.8486643491075\\
27	0.844483687576007\\
28	0.840244077564453\\
29	0.835957302170849\\
30	0.831633069624166\\
31	0.827279424693352\\
32	0.822903071270215\\
33	0.81850962684738\\
34	0.814103824473838\\
35	0.809689673974956\\
36	0.805270591403526\\
37	0.800849503580708\\
38	0.796428933001604\\
39	0.792011067183003\\
40	0.787597815621145\\
41	0.783190856832428\\
42	0.778791677416343\\
43	0.774401604668041\\
44	0.770021833948372\\
45	0.765653451770095\\
46	0.761297455363719\\
47	0.756954769332689\\
48	0.752626259886\\
49	0.748312747039538\\
50	0.744015015100219\\
51	0.739733821684924\\
52	0.735469905476253\\
53	0.731223992876638\\
54	0.726996803689491\\
55	0.722789055929319\\
56	0.718601469840876\\
57	0.71443477118967\\
58	0.710289693871641\\
59	0.706166981878157\\
60	0.702067390643105\\
61	0.697991687791525\\
62	0.693940653303662\\
63	0.689915079104274\\
64	0.685915768084403\\
65	0.681943532561467\\
66	0.677999192183355\\
67	0.674083571283096\\
68	0.670197495692666\\
69	0.666341789027326\\
70	0.662517268455744\\
71	0.658724739975716\\
72	0.654964993220719\\
73	0.651238795828535\\
74	0.647546887409814\\
75	0.643889973161527\\
76	0.640268717177664\\
77	0.636683735517191\\
78	0.633135589096904\\
79	0.629624776484445\\
80	0.626151726673982\\
81	0.622716791933833\\
82	0.619320240821483\\
83	0.615962251466619\\
84	0.612642905227042\\
85	0.609362180825154\\
86	0.606119949074229\\
87	0.602915968303493\\
88	0.599749880589116\\
89	0.596621208894435\\
90	0.593529355216949\\
91	0.590473599831827\\
92	0.587453101711829\\
93	0.584466900191667\\
94	0.581513917930986\\
95	0.578592965214461\\
96	0.575702745610059\\
97	0.572841862987582\\
98	0.570008829879303\\
99	0.567202077143205\\
100	0.564419964867191\\
};
\addlegendentry{Speed = 100km/h}

\addplot  [color=red!30!yellow, mark=diamond, mark options={ scale = 1.2}, mark repeat = 5]
  table[row sep=crcr]{%
1	0.26947405535904\\
2	0.453709942722607\\
3	0.580839538353395\\
4	0.669210278627581\\
5	0.730942186286974\\
6	0.774141303525321\\
7	0.804294514921989\\
8	0.825159297472147\\
9	0.839338026081785\\
10	0.848652929752137\\
11	0.854393619327133\\
12	0.85748225432905\\
13	0.85858490407264\\
14	0.858187391181101\\
15	0.856647452597437\\
16	0.854230954758315\\
17	0.851137270124105\\
18	0.847517218587988\\
19	0.843485862890509\\
20	0.839131711453958\\
21	0.834523391966041\\
22	0.829714529723641\\
23	0.82474734157885\\
24	0.819655303848128\\
25	0.814465147527032\\
26	0.809198361260489\\
27	0.803872331541515\\
28	0.798501213698162\\
29	0.793096601746665\\
30	0.787668046982267\\
31	0.782223462082562\\
32	0.776769438014903\\
33	0.771311494128369\\
34	0.765854276742494\\
35	0.760401717805342\\
36	0.754957162417696\\
37	0.749523471947592\\
38	0.744103107903204\\
39	0.738698200556883\\
40	0.73331060542087\\
41	0.727941949994026\\
42	0.722593672676107\\
43	0.717267055342803\\
44	0.711963250761929\\
45	0.706683305787294\\
46	0.701428181075774\\
47	0.696198767922571\\
48	0.690995902690617\\
49	0.685820379215308\\
50	0.680672959490053\\
51	0.675554382877282\\
52	0.670465374040538\\
53	0.665406649753466\\
54	0.66037892470928\\
55	0.655382916427891\\
56	0.650419349336435\\
57	0.645488958081404\\
58	0.64059249011632\\
59	0.635730707597426\\
60	0.630904388610646\\
61	0.626114327745984\\
62	0.621361336030053\\
63	0.616646240223705\\
64	0.611969881489269\\
65	0.607333113430878\\
66	0.602736799511511\\
67	0.598181809851667\\
68	0.593669017416967\\
69	0.589199293605436\\
70	0.584773503249526\\
71	0.5803924990533\\
72	0.57605711549127\\
73	0.571768162202274\\
74	0.567526416919334\\
75	0.563332617984522\\
76	0.559187456506409\\
77	0.555091568226492\\
78	0.551045525169995\\
79	0.547049827165351\\
80	0.543104893325398\\
81	0.539211053591622\\
82	0.535368540450422\\
83	0.531577480937144\\
84	0.527837889049355\\
85	0.524149658695151\\
86	0.520512557305168\\
87	0.516926220237969\\
88	0.51339014610763\\
89	0.50990369315926\\
90	0.506466076812863\\
91	0.50307636848819\\
92	0.499733495812943\\
93	0.496436244303885\\
94	0.49318326059502\\
95	0.489973057269155\\
96	0.486804019328874\\
97	0.483674412320444\\
98	0.480582392099662\\
99	0.47752601620237\\
100	0.474503256754675\\
};
\addlegendentry{Speed = 130km/h}

\end{axis}
\end{tikzpicture}%

%% file: Figures/P_NL_T.tex
\definecolor{mycolor1}{rgb}{0.00000,0.75000,0.75000}%
\begin{tikzpicture}
\pgfplotsset{
tick label style={font=\footnotesize},
label style={font=\footnotesize},
legend style={font=\scriptsize}
}
\tikzset{every mark/.append style={scale=1.2}}
\begin{axis}[%
width=\fwidth,
height=\fheight,
at={(0\fwidth,0\fheight)},
scale only axis,
xmin=0,
xmax=100,
xlabel style={font=\color{white!15!black}},
xlabel={$\text{Nodes per km (}\rho\text{)}$},
xlabel style={font=\color{white!15!black}},
ymin=0,
ymax=1,
xmajorgrids,
ymajorgrids,
ylabel style={font=\color{white!15!black}},
ylabel={$\text{P}_{\text{NL}}$},
axis background/.style={fill=white},
legend columns = 1,
legend pos =south east,
legend style={legend cell align=left, align=left, draw=white!15!black},
cycle list={%
{blue,mark=*, mark repeat = 5},
{red,mark=square, mark repeat = 5},
{black,mark=o, mark repeat = 5},
{brown!60!black,
mark options={fill=brown!40, scale = 1.2, mark repeat = 5},
mark=triangle*}}
]
\addplot
  table[row sep=crcr]{%
1	0.276805297486287\\
2	0.469231042693746\\
3	0.604719616335455\\
4	0.701289028161812\\
5	0.770921532328751\\
6	0.821684445204559\\
7	0.859074399599859\\
8	0.886879880397287\\
9	0.907741261584342\\
10	0.923518624037337\\
11	0.935536021367566\\
12	0.944745444486481\\
13	0.951838031374158\\
14	0.957320256578563\\
15	0.961566637826294\\
16	0.964856541818967\\
17	0.967400121043659\\
18	0.969356752831893\\
19	0.970848260222973\\
20	0.971968469833497\\
21	0.97279017696952\\
22	0.973370260683391\\
23	0.973753468388725\\
24	0.973975236445956\\
25	0.974063807082311\\
26	0.974041828031565\\
27	0.973927569283704\\
28	0.973735854522952\\
29	0.973478778585724\\
30	0.973166263426927\\
31	0.972806491464684\\
32	0.972406245267258\\
33	0.971971175294437\\
34	0.971506012064857\\
35	0.971014735163706\\
36	0.97050070855653\\
37	0.969966789465084\\
38	0.969415416396005\\
39	0.968848680651733\\
40	0.968268384692533\\
41	0.96767608998329\\
42	0.967073156393322\\
43	0.966460774780399\\
44	0.965839994050781\\
45	0.965211743722323\\
46	0.964576852810201\\
47	0.963936065691468\\
48	0.963290055475514\\
49	0.962639435304965\\
50	0.961984767929805\\
51	0.961326573832091\\
52	0.96066533812604\\
53	0.960001516415913\\
54	0.959335539759837\\
55	0.958667818859952\\
56	0.957998747576746\\
57	0.957328705847124\\
58	0.956658062070974\\
59	0.955987175018988\\
60	0.955316395304958\\
61	0.95464606645814\\
62	0.953976525625438\\
63	0.9533081039287\\
64	0.952641126499298\\
65	0.951975912210104\\
66	0.95131277312391\\
67	0.950652013677168\\
68	0.949993929618457\\
69	0.949338806722345\\
70	0.948686919301098\\
71	0.948038528538965\\
72	0.947393880676408\\
73	0.946753205074538\\
74	0.946116712193057\\
75	0.945484591518039\\
76	0.944857009478801\\
77	0.944234107395757\\
78	0.943615999503338\\
79	0.94300277109373\\
80	0.942394476828053\\
81	0.941791139261659\\
82	0.941192747629252\\
83	0.940599256933446\\
84	0.940010587377095\\
85	0.939426624175183\\
86	0.938847217776284\\
87	0.938272184516524\\
88	0.937701307720827\\
89	0.937134339257016\\
90	0.936571001538306\\
91	0.936010989959166\\
92	0.935453975738622\\
93	0.934899609134305\\
94	0.934347522980163\\
95	0.933797336491215\\
96	0.933248659270438\\
97	0.932701095446141\\
98	0.932154247863443\\
99	0.931607722250923\\
100	0.931061131283425\\
};
\addlegendentry{$\text{T}_{\text{RTO}} = \text{ 0.025s}$}

\addplot
  table[row sep=crcr]{%
1	0.274460337855998\\
2	0.46425546714595\\
3	0.597047446546264\\
4	0.690959983709364\\
5	0.758020119635939\\
6	0.806308318834657\\
7	0.841318761991507\\
8	0.86683054924447\\
9	0.885472306429741\\
10	0.899092306754741\\
11	0.90900380354887\\
12	0.916149400800132\\
13	0.921212328949368\\
14	0.92469253741733\\
15	0.926959236950407\\
16	0.928287523955524\\
17	0.928884143387317\\
18	0.928905772287815\\
19	0.928472107027739\\
20	0.927675309213348\\
21	0.926586878537515\\
22	0.925262692682906\\
23	0.9237467312206\\
24	0.922073847438614\\
25	0.920271846287639\\
26	0.918363052979548\\
27	0.916365505089785\\
28	0.914293864479763\\
29	0.912160119345587\\
30	0.90997412805456\\
31	0.907744042975639\\
32	0.905476642736477\\
33	0.903177594195234\\
34	0.900851660160566\\
35	0.89850286500492\\
36	0.896134627422345\\
37	0.893749867415876\\
38	0.891351092969021\\
39	0.88894047062207\\
40	0.886519883235098\\
41	0.884090977501744\\
42	0.881655203225962\\
43	0.879213845947781\\
44	0.876768054173219\\
45	0.874318862205439\\
46	0.871867209372026\\
47	0.869413956284013\\
48	0.866959898636356\\
49	0.8645057789595\\
50	0.862052296651793\\
51	0.859600116558521\\
52	0.857149876311821\\
53	0.85470219260411\\
54	0.852257666533962\\
55	0.849816888135965\\
56	0.847380440183795\\
57	0.844948901337593\\
58	0.842522848692\\
59	0.840102859769282\\
60	0.837689513992412\\
61	0.835283393665453\\
62	0.832885084482737\\
63	0.830495175584038\\
64	0.828114259169958\\
65	0.825742929689941\\
66	0.823381782614603\\
67	0.821031412804372\\
68	0.818692412487522\\
69	0.816365368862691\\
70	0.814050861343653\\
71	0.811749458467454\\
72	0.809461714490921\\
73	0.807188165704958\\
74	0.804929326500705\\
75	0.8026856852267\\
76	0.800457699881195\\
77	0.798245793688894\\
78	0.79605035061627\\
79	0.793871710884168\\
80	0.791710166540509\\
81	0.789565957159326\\
82	0.787439265734983\\
83	0.785330214842098\\
84	0.783238863132233\\
85	0.78116520223777\\
86	0.779109154151421\\
87	0.777070569146449\\
88	0.775049224297951\\
89	0.773044822659382\\
90	0.771056993141006\\
91	0.769085291128202\\
92	0.767129199867704\\
93	0.76518813263901\\
94	0.763261435716671\\
95	0.761348392117123\\
96	0.759448226111438\\
97	0.757560108473113\\
98	0.755683162418039\\
99	0.753816470182392\\
100	0.751959080173489\\
};
\addlegendentry{$\text{T}_{\text{RTO}}=\text{ 0.1s}$}

\addplot 
  table[row sep=crcr]{%
1	0.271341314755197\\
2	0.457653533545519\\
3	0.586892193226397\\
4	0.677321139991422\\
5	0.741025961812693\\
6	0.786103525428365\\
7	0.818043800865431\\
8	0.840612648178803\\
9	0.856422493938521\\
10	0.867305487637924\\
11	0.874560219454061\\
12	0.879116599715075\\
13	0.881647196930844\\
14	0.882643181424063\\
15	0.882466634263486\\
16	0.881386919369819\\
17	0.879606207349055\\
18	0.87727754682709\\
19	0.874517770354668\\
20	0.871416789045497\\
21	0.868044341256613\\
22	0.864454931698072\\
23	0.860691474166413\\
24	0.85678799839842\\
25	0.852771676236188\\
26	0.8486643491075\\
27	0.844483687576007\\
28	0.840244077564453\\
29	0.835957302170849\\
30	0.831633069624166\\
31	0.827279424693352\\
32	0.822903071270215\\
33	0.81850962684738\\
34	0.814103824473838\\
35	0.809689673974956\\
36	0.805270591403526\\
37	0.800849503580708\\
38	0.796428933001604\\
39	0.792011067183003\\
40	0.787597815621145\\
41	0.783190856832428\\
42	0.778791677416343\\
43	0.774401604668041\\
44	0.770021833948372\\
45	0.765653451770095\\
46	0.761297455363719\\
47	0.756954769332689\\
48	0.752626259886\\
49	0.748312747039538\\
50	0.744015015100219\\
51	0.739733821684924\\
52	0.735469905476253\\
53	0.731223992876638\\
54	0.726996803689491\\
55	0.722789055929319\\
56	0.718601469840876\\
57	0.71443477118967\\
58	0.710289693871641\\
59	0.706166981878157\\
60	0.702067390643105\\
61	0.697991687791525\\
62	0.693940653303662\\
63	0.689915079104274\\
64	0.685915768084403\\
65	0.681943532561467\\
66	0.677999192183355\\
67	0.674083571283096\\
68	0.670197495692666\\
69	0.666341789027326\\
70	0.662517268455744\\
71	0.658724739975716\\
72	0.654964993220719\\
73	0.651238795828535\\
74	0.647546887409814\\
75	0.643889973161527\\
76	0.640268717177664\\
77	0.636683735517191\\
78	0.633135589096904\\
79	0.629624776484445\\
80	0.626151726673982\\
81	0.622716791933833\\
82	0.619320240821483\\
83	0.615962251466619\\
84	0.612642905227042\\
85	0.609362180825154\\
86	0.606119949074229\\
87	0.602915968303493\\
88	0.599749880589116\\
89	0.596621208894435\\
90	0.593529355216949\\
91	0.590473599831827\\
92	0.587453101711829\\
93	0.584466900191667\\
94	0.581513917930986\\
95	0.578592965214461\\
96	0.575702745610059\\
97	0.572841862987582\\
98	0.570008829879303\\
99	0.567202077143205\\
100	0.564419964867191\\
};
\addlegendentry{$\text{T}_{\text{RTO}} = \text{ 0.2s}$}

\addplot
  table[row sep=crcr]{%
1	0.262036061105739\\
2	0.438066377620407\\
3	0.556929294413567\\
4	0.637302189840702\\
5	0.691436991424122\\
6	0.727470892389412\\
7	0.750873743213571\\
8	0.76536573583997\\
9	0.773506113137598\\
10	0.77707515713438\\
11	0.777324793532707\\
12	0.775144770052859\\
13	0.771174001548313\\
14	0.765875931844751\\
15	0.759590044193285\\
16	0.75256740723044\\
17	0.744995433844309\\
18	0.737015283947711\\
19	0.728734205797154\\
20	0.720234364242573\\
21	0.711579209826242\\
22	0.702818112160452\\
23	0.693989758252146\\
24	0.685124665056541\\
25	0.676247051837166\\
26	0.667376246314201\\
27	0.658527748779448\\
28	0.649714043454867\\
29	0.640945221736813\\
30	0.632229464456983\\
31	0.623573417758077\\
32	0.614982488151421\\
33	0.606461075773853\\
34	0.598012760079732\\
35	0.589640448691413\\
36	0.581346497535115\\
37	0.57313280845816\\
38	0.565000909078777\\
39	0.556952018532401\\
40	0.548987101955156\\
41	0.541106915918511\\
42	0.533312046549043\\
43	0.525602941697618\\
44	0.51797993823594\\
45	0.510443285335366\\
46	0.502993164408212\\
47	0.495629706254083\\
48	0.488353005844789\\
49	0.481163135094519\\
50	0.474060153892428\\
51	0.4670441196188\\
52	0.460115095320671\\
53	0.453273156686037\\
54	0.446518397925805\\
55	0.439850936648155\\
56	0.433270917790042\\
57	0.426778516654243\\
58	0.420373941087163\\
59	0.414057432822072\\
60	0.407829268004095\\
61	0.401689756906979\\
62	0.395639242847111\\
63	0.389678100297319\\
64	0.383806732201569\\
65	0.378025566491619\\
66	0.372335051808006\\
67	0.366735652430309\\
68	0.361227842425446\\
69	0.355812099027713\\
70	0.350488895270438\\
71	0.34525869189628\\
72	0.340121928581428\\
73	0.335079014518141\\
74	0.330130318410028\\
75	0.325276157945272\\
76	0.320516788824337\\
77	0.315852393430514\\
78	0.311283069243801\\
79	0.306808817110755\\
80	0.302429529495022\\
81	0.298144978844879\\
82	0.293954806225097\\
83	0.28985851037042\\
84	0.285855437326712\\
85	0.281944770852896\\
86	0.278125523762015\\
87	0.274396530382609\\
88	0.270756440321903\\
89	0.267203713709591\\
90	0.263736618095132\\
91	0.26035322716202\\
92	0.257051421409337\\
93	0.253828890933802\\
94	0.250683140424445\\
95	0.247611496456816\\
96	0.244611117144531\\
97	0.241679004172874\\
98	0.238812017202622\\
99	0.236006890592483\\
100	0.233260252346071\\
};
\addlegendentry{$\text{T}_{\text{RTO}}=\text{ 0.5s}$}

\addplot  [color=red!30!yellow, mark=diamond, mark options={ scale = 1.2}, mark repeat = 5]
  table[row sep=crcr]{%
1	0.246698591844485\\
2	0.406137909832417\\
3	0.508626061333733\\
4	0.573498145665648\\
5	0.613245010246332\\
6	0.636035233162007\\
7	0.647274766325337\\
8	0.650583035609238\\
9	0.648410937034109\\
10	0.642436519341874\\
11	0.633821099935908\\
12	0.623376807561275\\
13	0.611677320157804\\
14	0.599131798712811\\
15	0.586034738300588\\
16	0.572599907777055\\
17	0.558983677674115\\
18	0.545301205452944\\
19	0.531637769757451\\
20	0.518056780894349\\
21	0.504605494138346\\
22	0.491319121771699\\
23	0.478223819502035\\
24	0.465338874990846\\
25	0.452678326109317\\
26	0.440252168245805\\
27	0.428067263049908\\
28	0.416128028493899\\
29	0.40443696745894\\
30	0.392995076123239\\
31	0.381802162157046\\
32	0.370857094696867\\
33	0.360158002307537\\
34	0.349702430976362\\
35	0.33948747115401\\
36	0.329509860637789\\
37	0.3197660684565\\
38	0.310252363701015\\
39	0.300964872336404\\
40	0.291899624347761\\
41	0.283052593053799\\
42	0.274419728026924\\
43	0.265996982754766\\
44	0.257780337943149\\
45	0.249765821177426\\
46	0.241949523515491\\
47	0.234327613472373\\
48	0.226896348766139\\
49	0.219652086122613\\
50	0.212591289378311\\
51	0.205710536073839\\
52	0.199006522691683\\
53	0.192476068660894\\
54	0.186116119225486\\
55	0.179923747252174\\
56	0.173896154035813\\
57	0.168030669146693\\
58	0.16232474935246\\
59	0.156775976638306\\
60	0.151382055342077\\
61	0.14614080841576\\
62	0.141050172821339\\
63	0.136108194067095\\
64	0.131313019890016\\
65	0.126662893090926\\
66	0.122156143531319\\
67	0.117791179304514\\
68	0.113566477098648\\
69	0.109480571775222\\
70	0.105532045194201\\
71	0.101719514325184\\
72	0.0980416186936718\\
73	0.0944970072220088\\
74	0.0910843245359781\\
75	0.087802196820223\\
76	0.0846492173184659\\
77	0.0816239315877492\\
78	0.0787248226294223\\
79	0.0759502960331065\\
80	0.0732986652831379\\
81	0.070768137389709\\
82	0.068356799018794\\
83	0.066062603305591\\
84	0.0638833575452797\\
85	0.0618167119619876\\
86	0.059860149761559\\
87	0.0580109786756423\\
88	0.0562663242033227\\
89	0.0546231247516523\\
90	0.0530781288675901\\
91	0.05162789474074\\
92	0.0502687921385949\\
93	0.0489970069135493\\
94	0.0478085481936356\\
95	0.0466992583367431\\
96	0.0456648256911263\\
97	0.0447008001635255\\
98	0.0438026115506391\\
99	0.0429655905405304\\
100	0.0421849922385787\\
};
\addlegendentry{$\text{T}_{\text{RTO}}=\text{ 1s}$}

\end{axis}
\end{tikzpicture}%

%% file: Figures/T_MIMO.tex
\tikzstyle{every pin}=[fill=white,
draw=black,
font=\footnotesize]
\begin{tikzpicture}
\pgfplotsset{
tick label style={font=\footnotesize},
label style={font=\footnotesize},
legend style={font=\scriptsize}
}
\tikzset{every mark/.append style={scale=1.2}}
\begin{axis}[%
width=\fwidth,
height=\fheight,
at={(0\fwidth,0\fheight)},
scale only axis,
xmin=0,
xmax=100,
xlabel style={font=\color{white!15!black}},
xlabel={$\text{Nodes per km (}\rho\text{)}$},
xlabel style={font=\color{white!15!black}},
ymin=0,
ymax=1,
xmajorgrids,
ymajorgrids,
ylabel style={font=\color{white!15!black}},
ylabel={Communication Duration [Ratio]},
axis background/.style={fill=white},
legend columns = 1,
legend pos =south east,
legend style={legend cell align=left, align=left, draw=white!15!black},
cycle list={%
{blue,mark=*, mark repeat = 5},
{red,mark=square, mark repeat = 5},
{black,mark=o, mark repeat = 5},
{brown!60!black,
mark options={fill=brown!40, scale = 1.2, mark repeat = 5},
mark=triangle*}}
]
\addplot 
  table[row sep=crcr]{%
1	0.184174484785114\\
2	0.32461056144542\\
3	0.432818619538242\\
4	0.517054228933617\\
5	0.583273046098713\\
6	0.635806738424154\\
7	0.677835372179812\\
8	0.711716694411117\\
9	0.73921630907019\\
10	0.761669516434478\\
11	0.780095962012577\\
12	0.795281541508037\\
13	0.807837433817705\\
14	0.818243033665205\\
15	0.826877456150692\\
16	0.834042859929845\\
17	0.839981862719851\\
18	0.844890654395526\\
19	0.848928950400349\\
20	0.852227605662835\\
21	0.854894482519444\\
22	0.857019005548692\\
23	0.85867572154625\\
24	0.859927100346961\\
25	0.860825752359227\\
26	0.861416194961283\\
27	0.861736267741451\\
28	0.86181827272823\\
29	0.861689897973052\\
30	0.861374969494145\\
31	0.860894066497396\\
32	0.860265027115653\\
33	0.859503366037689\\
34	0.858622620881827\\
35	0.857634640675675\\
36	0.856549827086291\\
37	0.855377336920967\\
38	0.854125252749988\\
39	0.852800727185147\\
40	0.85141010530287\\
41	0.849959028868094\\
42	0.848452525348891\\
43	0.846895084176301\\
44	0.845290722271899\\
45	0.84364304051558\\
46	0.841955272541517\\
47	0.840230327017887\\
48	0.838470824375776\\
49	0.836679128796195\\
50	0.834857376135239\\
51	0.8330074983606\\
52	0.831131244984156\\
53	0.829230201901519\\
54	0.827305807987874\\
55	0.825359369747855\\
56	0.823392074273871\\
57	0.821405000730805\\
58	0.819399130554244\\
59	0.817375356523254\\
60	0.815334490846669\\
61	0.813277272382964\\
62	0.811204373097753\\
63	0.809116403849267\\
64	0.80701391958035\\
65	0.804897423985483\\
66	0.802767373712616\\
67	0.800624182152149\\
68	0.798468222858904\\
69	0.796299832647358\\
70	0.794119314395549\\
71	0.791926939588849\\
72	0.789722950631137\\
73	0.7875075629477\\
74	0.785280966901409\\
75	0.783043329541251\\
76	0.780794796200179\\
77	0.77853549195736\\
78	0.776265522978225\\
79	0.773984977744305\\
80	0.77169392818352\\
81	0.769392430710496\\
82	0.767080527185425\\
83	0.764758245799167\\
84	0.762425601891441\\
85	0.760082598708294\\
86	0.757729228104408\\
87	0.755365471195235\\
88	0.752991298963479\\
89	0.750606672823993\\
90	0.748211545150768\\
91	0.745805859769337\\
92	0.743389552417602\\
93	0.740962551177814\\
94	0.738524776882184\\
95	0.73607614349436\\
96	0.733616558468812\\
97	0.731145923089991\\
98	0.728664132792932\\
99	0.726171077466854\\
100	0.723666641743153\\
};
\addlegendentry{4x4}

\addplot
  table[row sep=crcr]{%
1	0.228537703906177\\
2	0.391972806451515\\
3	0.510620121548375\\
4	0.598018897705361\\
5	0.663285262393139\\
6	0.71263667981856\\
7	0.750373282628639\\
8	0.779511803143575\\
9	0.802198612089401\\
10	0.81998183986631\\
11	0.833992876531232\\
12	0.845068982685357\\
13	0.853837183525068\\
14	0.860772397201866\\
15	0.866238205179892\\
16	0.870515786508201\\
17	0.873824685522363\\
18	0.876337880100423\\
19	0.878192828251873\\
20	0.879499646757799\\
21	0.880347223774448\\
22	0.880807828578389\\
23	0.880940617922424\\
24	0.880794325067637\\
25	0.880409338230564\\
26	0.879819319171542\\
27	0.879052472740742\\
28	0.878132549514664\\
29	0.8770796428682\\
30	0.875910826640989\\
31	0.874640668377159\\
32	0.873281644826456\\
33	0.871844480202522\\
34	0.870338423037608\\
35	0.868771473948702\\
36	0.867150573945739\\
37	0.865481760855772\\
38	0.863770299851769\\
39	0.862020792846173\\
40	0.860237270552117\\
41	0.858423270265329\\
42	0.856581901829642\\
43	0.854715903782022\\
44	0.852827691302051\\
45	0.850919397294482\\
46	0.848992907695935\\
47	0.847049891905399\\
48	0.845091829083444\\
49	0.843120030939284\\
50	0.84113566152238\\
51	0.839139754451216\\
52	0.837133227943027\\
53	0.835116897951193\\
54	0.833091489670002\\
55	0.83105764762721\\
56	0.829015944552215\\
57	0.826966889180274\\
58	0.824910933130216\\
59	0.82284847697378\\
60	0.820779875598304\\
61	0.818705442950643\\
62	0.816625456238355\\
63	0.81454015965418\\
64	0.812449767681143\\
65	0.810354468028308\\
66	0.808254424240776\\
67	0.806149778022106\\
68	0.804040651302545\\
69	0.801927148082398\\
70	0.79980935607627\\
71	0.797687348180846\\
72	0.795561183786153\\
73	0.793430909947899\\
74	0.79129656243644\\
75	0.789158166676073\\
76	0.78701573858684\\
77	0.784869285339569\\
78	0.782718806033698\\
79	0.78056429230633\\
80	0.778405728880025\\
81	0.776243094055993\\
82	0.774076360158615\\
83	0.771905493936542\\
84	0.769730456925094\\
85	0.767551205774086\\
86	0.765367692544847\\
87	0.763179864979699\\
88	0.760987666746885\\
89	0.758791037663558\\
90	0.756589913899187\\
91	0.754384228161488\\
92	0.752173909866735\\
93	0.749958885296138\\
94	0.74773907773978\\
95	0.74551440762945\\
96	0.743284792661557\\
97	0.741050147911211\\
98	0.738810385938408\\
99	0.736565416887192\\
100	0.734315148578542\\
};
\addlegendentry{4x16}

\addplot
  table[row sep=crcr]{%
1	0.261960160236534\\
2	0.444180837357088\\
3	0.572543540417961\\
4	0.664135018513422\\
5	0.730300110550116\\
6	0.778644856567908\\
7	0.814328590164017\\
8	0.840895184216868\\
9	0.860810240649716\\
10	0.875811123308389\\
11	0.887137374691543\\
12	0.895684090716514\\
13	0.902105195028152\\
14	0.906883786446233\\
15	0.910380607072099\\
16	0.912867809852293\\
17	0.914552739878375\\
18	0.915594858470337\\
19	0.916117909058078\\
20	0.916218747671371\\
21	0.91597381235126\\
22	0.915443905301994\\
23	0.914677758268985\\
24	0.913714712706862\\
25	0.912586750496371\\
26	0.911320044296475\\
27	0.909936149811814\\
28	0.908452929116785\\
29	0.906885270522712\\
30	0.90524565345585\\
31	0.903544594476127\\
32	0.901791001556083\\
33	0.899992457112256\\
34	0.898155445373365\\
35	0.896285536010953\\
36	0.894387533213043\\
37	0.892465597309198\\
38	0.890523344481502\\
39	0.888563928893879\\
40	0.886590110648789\\
41	0.884604312267102\\
42	0.88260866583337\\
43	0.880605052516562\\
44	0.878595135837612\\
45	0.876580389788124\\
46	0.874562122693327\\
47	0.872541497544321\\
48	0.87051954939055\\
49	0.868497200275852\\
50	0.86647527211486\\
51	0.864454497836503\\
52	0.862435531064668\\
53	0.860418954559796\\
54	0.858405287607484\\
55	0.856394992509229\\
56	0.854388480305019\\
57	0.852386115836502\\
58	0.850388222242155\\
59	0.84839508496148\\
60	0.846406955313341\\
61	0.844424053703579\\
62	0.842446572508745\\
63	0.840474678675855\\
64	0.838508516072222\\
65	0.836548207614521\\
66	0.834593857202141\\
67	0.832645551476389\\
68	0.830703361424163\\
69	0.828767343842253\\
70	0.826837542676288\\
71	0.824913990246601\\
72	0.822996708371705\\
73	0.821085709398827\\
74	0.819180997149816\\
75	0.817282567789768\\
76	0.815390410624946\\
77	0.813504508835831\\
78	0.811624840150552\\
79	0.809751377463401\\
80	0.807884089402724\\
81	0.806022940852038\\
82	0.804167893427896\\
83	0.802318905917723\\
84	0.800475934680577\\
85	0.798638934013526\\
86	0.796807856486167\\
87	0.794982653245568\\
88	0.793163274293796\\
89	0.791349668739991\\
90	0.789541785028866\\
91	0.78773957114731\\
92	0.785942974810732\\
93	0.784151943630614\\
94	0.782366425264696\\
95	0.780586367551066\\
96	0.778811718627409\\
97	0.777042427036528\\
98	0.775278441819225\\
99	0.773519712595523\\
100	0.771766189635181\\
};
\addlegendentry{64x4}

\addplot 
  table[row sep=crcr]{%
1	0.270059701424074\\
2	0.458538123841035\\
3	0.591146899570919\\
4	0.685282073639983\\
5	0.752704902175927\\
6	0.801402208224524\\
7	0.836836874263349\\
8	0.86277810833129\\
9	0.881851308956442\\
10	0.895903229338441\\
11	0.906245867184076\\
12	0.913820558672197\\
13	0.919309300198555\\
14	0.923210931130286\\
15	0.925893736137974\\
16	0.92763209187431\\
17	0.928632224877132\\
18	0.929050474502552\\
19	0.929006352728296\\
20	0.92859196135696\\
21	0.927878837991218\\
22	0.926922972343449\\
23	0.925768510288457\\
24	0.924450509511822\\
25	0.922997004590118\\
26	0.921430565582762\\
27	0.919769482517432\\
28	0.918028671648912\\
29	0.916220373414426\\
30	0.914354693421932\\
31	0.912440024408678\\
32	0.910483377383866\\
33	0.908490643067991\\
34	0.906466799522624\\
35	0.904416078005644\\
36	0.902342096216886\\
37	0.900247965952002\\
38	0.898136380567015\\
39	0.896009686433913\\
40	0.893869941638219\\
41	0.891718964458602\\
42	0.889558373622391\\
43	0.887389621908658\\
44	0.885214024342853\\
45	0.883032781971158\\
46	0.880847002002123\\
47	0.878657714945059\\
48	0.876465889249431\\
49	0.874272443849871\\
50	0.872078258941741\\
51	0.869884185248191\\
52	0.86769105198806\\
53	0.865499673712095\\
54	0.863310856141064\\
55	0.861125401111572\\
56	0.858944110712868\\
57	0.856767790679453\\
58	0.854597253089351\\
59	0.852433318405777\\
60	0.850276816890268\\
61	0.848128589407789\\
62	0.845989487638551\\
63	0.843860373707216\\
64	0.841742119237475\\
65	0.839635603838743\\
66	0.837541713031611\\
67	0.835461335619857\\
68	0.833395360518963\\
69	0.831344673054338\\
70	0.82931015074654\\
71	0.827292658605866\\
72	0.825293043964403\\
73	0.823312130880127\\
74	0.821350714154681\\
75	0.819409553013849\\
76	0.817489364507578\\
77	0.815590816694193\\
78	0.813714521681264\\
79	0.81186102860316\\
80	0.810030816622352\\
81	0.808224288047979\\
82	0.806441761670642\\
83	0.804683466416795\\
84	0.802949535429143\\
85	0.801240000680974\\
86	0.799554788232154\\
87	0.79789371423249\\
88	0.796256481774149\\
89	0.794642678688754\\
90	0.793051776376665\\
91	0.791483129745762\\
92	0.789935978324871\\
93	0.788409448602955\\
94	0.786902557629442\\
95	0.785414217893887\\
96	0.783943243484722\\
97	0.782488357507515\\
98	0.781048200723221\\
99	0.779621341346694\\
100	0.778206285925572\\
};
\addlegendentry{64x16}

\node[pin=290:Saturation,draw=black]
at (axis cs:22,0.85) {};

\end{axis}
\end{tikzpicture}%

%% file: Figures/T_V.tex
\definecolor{mycolor1}{rgb}{0.00000,0.75000,0.75000}%
\begin{tikzpicture}
\pgfplotsset{
tick label style={font=\footnotesize},
label style={font=\footnotesize},
legend style={font=\scriptsize}
}
\tikzset{every mark/.append style={scale=1.2}}
\begin{axis}[%
width=\fwidth,
height=\fheight,
at={(0\fwidth,0\fheight)},
scale only axis,
xmin=0,
xmax=100,
xlabel style={font=\color{white!15!black}},
xlabel={$\text{Nodes per km (}\rho\text{)}$},
xlabel style={font=\color{white!15!black}},
ymin=0,
ymax=1,
xmajorgrids,
ymajorgrids,
ylabel style={font=\color{white!15!black}},
ylabel={Communication Duration [Ratio]},
axis background/.style={fill=white},
legend columns = 1,
legend pos =south east,
legend style={legend cell align=left, align=left, draw=white!15!black},
cycle list={%
{blue,mark=*, mark repeat = 5},
{red,mark=square, mark repeat = 5},
{black,mark=o, mark repeat = 5},
{brown!60!black,
mark options={fill=brown!40, scale = 1.2, mark repeat = 5},
mark=triangle*}}
]
\addplot 
  table[row sep=crcr]{%
1	0.276749109352079\\
2	0.46951418615096\\
3	0.605481702080708\\
4	0.702569308898962\\
5	0.772721240040196\\
6	0.823991798630252\\
7	0.861874482381789\\
8	0.890158351116858\\
9	0.911485560675626\\
10	0.92771818983761\\
11	0.94018214046739\\
12	0.949830992069307\\
13	0.95735720245323\\
14	0.963268324347037\\
15	0.967939748762394\\
16	0.971651546789116\\
17	0.974614438064506\\
18	0.976988256372253\\
19	0.978895192229826\\
20	0.980429368318887\\
21	0.98166381870138\\
22	0.982655615167497\\
23	0.983449660877776\\
24	0.98408151816464\\
25	0.984579531223056\\
26	0.984966430369194\\
27	0.985260552491167\\
28	0.985476775456146\\
29	0.985627237953169\\
30	0.98572189737719\\
31	0.985768964717574\\
32	0.985775245489047\\
33	0.985746408475963\\
34	0.985687198708737\\
35	0.985601607124149\\
36	0.98549300640539\\
37	0.985364260281872\\
38	0.985217811898955\\
39	0.985055755602358\\
40	0.984879895519083\\
41	0.984691793578261\\
42	0.984492809048715\\
43	0.984284131230997\\
44	0.984066806601153\\
45	0.983841761437923\\
46	0.983609820756408\\
47	0.983371724207622\\
48	0.983128139473438\\
49	0.982879673583572\\
50	0.982626882499283\\
51	0.982370279242481\\
52	0.982110340796478\\
53	0.981847513961785\\
54	0.981582220316077\\
55	0.981314860399696\\
56	0.981045817225057\\
57	0.980775459190446\\
58	0.980504142463394\\
59	0.980232212887076\\
60	0.979960007453484\\
61	0.97968785537945\\
62	0.979416078815797\\
63	0.979144993215477\\
64	0.978874907383239\\
65	0.978606123227621\\
66	0.978338935234655\\
67	0.978073629682818\\
68	0.977810483619189\\
69	0.977549763617968\\
70	0.977291724344507\\
71	0.977036606950122\\
72	0.976784637325706\\
73	0.976536024245048\\
74	0.97629095743175\\
75	0.976049605586725\\
76	0.975812114416032\\
77	0.975578604701464\\
78	0.97534917045843\\
79	0.975123877227063\\
80	0.974902760543367\\
81	0.974685824637025\\
82	0.974473041401094\\
83	0.974264349676763\\
84	0.97405965489237\\
85	0.973858829091185\\
86	0.973661711376147\\
87	0.973468108792202\\
88	0.973277797658283\\
89	0.973090525351164\\
90	0.972906012532957\\
91	0.972723955802916\\
92	0.972544030742794\\
93	0.972365895313956\\
94	0.972189193553558\\
95	0.972013559507545\\
96	0.971838621329729\\
97	0.971664005469613\\
98	0.971489340867152\\
99	0.971314263070595\\
100	0.971138418194128\\
};
\addlegendentry{Speed = 10km/h}

\addplot 
  table[row sep=crcr]{%
1	0.275910794315033\\
2	0.468135972620849\\
3	0.603678877335352\\
4	0.700392095743982\\
5	0.770196939323345\\
6	0.82113924176817\\
7	0.858708673696917\\
8	0.886691831621914\\
9	0.907728798833073\\
10	0.923679756789146\\
11	0.93586888126573\\
12	0.94524822682083\\
13	0.952508943720869\\
14	0.958157491738228\\
15	0.962568367204484\\
16	0.96602092148776\\
17	0.968725303948479\\
18	0.970840902777925\\
19	0.972489565121515\\
20	0.973765152972574\\
21	0.974740505871416\\
22	0.975472553564361\\
23	0.976006098464398\\
24	0.976376634419467\\
25	0.976612462168722\\
26	0.976736287849475\\
27	0.976766438904583\\
28	0.976717794922888\\
29	0.976602504700184\\
30	0.976430541969125\\
31	0.976210138634075\\
32	0.975948124446044\\
33	0.975650194806925\\
34	0.975321123055722\\
35	0.974964929636601\\
36	0.974585017602898\\
37	0.974184281704035\\
38	0.973765196639356\\
39	0.973329888802922\\
40	0.972880194884161\\
41	0.972417709954984\\
42	0.971943827109283\\
43	0.971459770284215\\
44	0.970966621553627\\
45	0.970465343919567\\
46	0.969956800420386\\
47	0.96944177021077\\
48	0.968920962139878\\
49	0.968395026251238\\
50	0.967864563546236\\
51	0.967330134287522\\
52	0.966792265066021\\
53	0.966251454812723\\
54	0.965708179902038\\
55	0.965162898465603\\
56	0.96461605401285\\
57	0.964068078436135\\
58	0.963519394463351\\
59	0.962970417608922\\
60	0.962421557664326\\
61	0.961873219761774\\
62	0.961325805038659\\
63	0.960779710926073\\
64	0.96023533108144\\
65	0.959693054983507\\
66	0.959153267206888\\
67	0.95861634639337\\
68	0.958082663937985\\
69	0.957552582409311\\
70	0.957026453725716\\
71	0.956504617111788\\
72	0.955987396862387\\
73	0.955475099945159\\
74	0.954968013475919\\
75	0.954466402105055\\
76	0.953970505356672\\
77	0.953480534965592\\
78	0.952996672260292\\
79	0.952519065642388\\
80	0.952047828214922\\
81	0.951583035612651\\
82	0.951124724087393\\
83	0.950672888900167\\
84	0.950227483069415\\
85	0.949788416520683\\
86	0.949355555678006\\
87	0.948928723530619\\
88	0.948507700200914\\
89	0.948092224030423\\
90	0.947681993190753\\
91	0.947276667815522\\
92	0.94687587263809\\
93	0.946479200108524\\
94	0.946086213951849\\
95	0.945696453119018\\
96	0.945309436072304\\
97	0.944924665338377\\
98	0.944541632255677\\
99	0.94415982183798\\
100	0.943778717673537\\
};
\addlegendentry{Speed = 20km/h}

\addplot 
  table[row sep=crcr]{%
1	0.275073091372732\\
2	0.466759537137921\\
3	0.601879200251726\\
4	0.698219564500792\\
5	0.767679033968843\\
6	0.818294983274765\\
7	0.855553256531295\\
8	0.88323798223735\\
9	0.903987165248881\\
10	0.919659084853579\\
11	0.931576187714836\\
12	0.940689002267278\\
13	0.94768737159068\\
14	0.953076663992724\\
15	0.957230483375052\\
16	0.960427464086904\\
17	0.962877188135597\\
18	0.964738601099011\\
19	0.966133209673519\\
20	0.967154617959982\\
21	0.967875473604823\\
22	0.96835256674981\\
23	0.96863060130524\\
24	0.968745004687902\\
25	0.968724036051555\\
26	0.968590379053897\\
27	0.968362353233585\\
28	0.968054841297336\\
29	0.967680003411764\\
30	0.967247830790843\\
31	0.966766577287398\\
32	0.966243097821169\\
33	0.965683115250272\\
34	0.965091431973281\\
35	0.964472098609418\\
36	0.963828549169526\\
37	0.96316370993182\\
38	0.962480087580209\\
39	0.961779840908513\\
40	0.96106483943897\\
41	0.960336711572456\\
42	0.959596884325825\\
43	0.958846616277358\\
44	0.958087025003837\\
45	0.957319110029564\\
46	0.956543772101285\\
47	0.955761829440394\\
48	0.95497403149529\\
49	0.954181070614588\\
50	0.953383591980399\\
51	0.952582202075518\\
52	0.951777475905957\\
53	0.950969963157665\\
54	0.950160193432097\\
55	0.949348680677303\\
56	0.948535926908579\\
57	0.947722425294262\\
58	0.946908662667328\\
59	0.946095121511285\\
60	0.945282281459249\\
61	0.94447062033738\\
62	0.943660614777949\\
63	0.94285274042297\\
64	0.942047471736145\\
65	0.941245281439149\\
66	0.940446639587292\\
67	0.939652012299892\\
68	0.938861860161544\\
69	0.93807663631235\\
70	0.937296784247548\\
71	0.936522735350052\\
72	0.935754906182936\\
73	0.934993695572826\\
74	0.934239481519281\\
75	0.933492617969661\\
76	0.932753431503155\\
77	0.932022217971901\\
78	0.931299239150884\\
79	0.930584719451639\\
80	0.929878842757523\\
81	0.929181749440049\\
82	0.928493533616822\\
83	0.927814240711173\\
84	0.927143865372245\\
85	0.926482349811332\\
86	0.925829582606098\\
87	0.925185398018573\\
88	0.92454957586583\\
89	0.923921841973838\\
90	0.923301869235466\\
91	0.922689279283095\\
92	0.922083644775021\\
93	0.921484492283094\\
94	0.920891305757196\\
95	0.920303530530514\\
96	0.919720577818473\\
97	0.919141829654134\\
98	0.918566644193947\\
99	0.9179943613206\\
100	0.91742430846421\\
};
\addlegendentry{Speed = 30km/h}

\addplot
  table[row sep=crcr]{%
1	0.269226270010446\\
2	0.457174065072578\\
3	0.58936908684978\\
4	0.683142004614403\\
5	0.750231264366396\\
6	0.798615295332723\\
7	0.833753150199068\\
8	0.859411529347692\\
9	0.878213705695854\\
10	0.892004490814542\\
11	0.902094126841523\\
12	0.909422410303815\\
13	0.914670027495374\\
14	0.91833473114857\\
15	0.920783923434728\\
16	0.922291276876314\\
17	0.923062464899221\\
18	0.923253398707182\\
19	0.922983263754648\\
20	0.92234391687206\\
21	0.921406715418205\\
22	0.920227519758967\\
23	0.918850386106493\\
24	0.91731031317266\\
25	0.915635300095524\\
26	0.913847899382712\\
27	0.91196639696591\\
28	0.910005715005823\\
29	0.907978107172994\\
30	0.905893697580382\\
31	0.90376090117568\\
32	0.901586753703742\\
33	0.899377172269223\\
34	0.897137162327705\\
35	0.894870983088467\\
36	0.892582280452829\\
37	0.890274194473478\\
38	0.88794944671165\\
39	0.885610411652373\\
40	0.883259175412675\\
41	0.880897584270305\\
42	0.878527284996745\\
43	0.876149758558279\\
44	0.873766348422735\\
45	0.871378284454932\\
46	0.868986703184274\\
47	0.866592665070513\\
48	0.864197169269091\\
49	0.861801166298236\\
50	0.859405568930638\\
51	0.857011261568778\\
52	0.854619108311518\\
53	0.852229959877884\\
54	0.849844659520073\\
55	0.847464048030077\\
56	0.845088967921771\\
57	0.842720266851924\\
58	0.840358800328617\\
59	0.838005433743472\\
60	0.835661043754458\\
61	0.833326519038495\\
62	0.831002760427411\\
63	0.828690680436718\\
64	0.826391202194178\\
65	0.824105257773892\\
66	0.821833785941755\\
67	0.819577729319371\\
68	0.817338030975922\\
69	0.815115630460878\\
70	0.812911459294877\\
71	0.810726435941351\\
72	0.808561460287625\\
73	0.806417407671012\\
74	0.80429512249283\\
75	0.802195411471195\\
76	0.800119036591601\\
77	0.79806670782267\\
78	0.796039075672785\\
79	0.79403672367138\\
80	0.792060160866323\\
81	0.790109814435786\\
82	0.788186022519061\\
83	0.786289027375727\\
84	0.784418968986163\\
85	0.782575879208412\\
86	0.780759676606695\\
87	0.778970162065185\\
88	0.777207015296981\\
89	0.775469792352323\\
90	0.773757924222056\\
91	0.772070716622078\\
92	0.770407351032057\\
93	0.768766887047283\\
94	0.767148266086085\\
95	0.765550316477205\\
96	0.763971759932002\\
97	0.76241121938564\\
98	0.76086722816993\\
99	0.759338240458453\\
100	0.75782264290244\\
};
\addlegendentry{Speed = 100km/h}

\addplot [color=red!30!yellow, mark=diamond, mark options={ scale = 1.2}, mark repeat = 5]
  table[row sep=crcr]{%
1	0.266729623981826\\
2	0.453092428129502\\
3	0.584054221533123\\
4	0.676749313935427\\
5	0.742847799826469\\
6	0.790302975952165\\
7	0.824562395737402\\
8	0.849385196280452\\
9	0.867388227656067\\
10	0.88041043766566\\
11	0.889756774237143\\
12	0.896362398552542\\
13	0.900904059212947\\
14	0.9038762481389\\
15	0.905643723289313\\
16	0.906478050443299\\
17	0.906583250343987\\
18	0.90611395645178\\
19	0.905188380805935\\
20	0.903897650538528\\
21	0.90231258635452\\
22	0.900488663433487\\
23	0.898469670622822\\
24	0.896290430150815\\
25	0.893978834165074\\
26	0.891557380818338\\
27	0.889044341123053\\
28	0.886454651486254\\
29	0.883800601053159\\
30	0.881092364551382\\
31	0.878338418054592\\
32	0.875545865465011\\
33	0.872720696497444\\
34	0.869867991796978\\
35	0.866992087018323\\
36	0.864096704868208\\
37	0.86118506199962\\
38	0.858259956058589\\
39	0.855323836983718\\
40	0.852378865746067\\
41	0.849426963019596\\
42	0.846469849736458\\
43	0.84350908106751\\
44	0.840546075047036\\
45	0.837582136809817\\
46	0.834618479211911\\
47	0.831656240451351\\
48	0.82869649918201\\
49	0.825740287515941\\
50	0.822788602231117\\
51	0.819842414438465\\
52	0.816902677911147\\
53	0.813970336237755\\
54	0.81104632892745\\
55	0.808131596567655\\
56	0.805227085112512\\
57	0.802333749362009\\
58	0.799452555676847\\
59	0.796584483962117\\
60	0.793730528943321\\
61	0.790891700750884\\
62	0.788069024823785\\
63	0.785263541139075\\
64	0.782476302771795\\
65	0.779708373788925\\
66	0.776960826481434\\
67	0.774234737940275\\
68	0.77153118598503\\
69	0.768851244457973\\
70	0.766195977901414\\
71	0.763566435642251\\
72	0.760963645314631\\
73	0.75838860585944\\
74	0.755842280047735\\
75	0.753325586584323\\
76	0.750839391857051\\
77	0.748384501407048\\
78	0.745961651204793\\
79	0.743571498826406\\
80	0.741214614633608\\
81	0.73889147306922\\
82	0.736602444187583\\
83	0.73434778554562\\
84	0.73212763458519\\
85	0.729942001640619\\
86	0.727790763706678\\
87	0.725673659101467\\
88	0.723590283155651\\
89	0.72154008505394\\
90	0.719522365946707\\
91	0.717536278438917\\
92	0.715580827550337\\
93	0.713654873225117\\
94	0.711757134450605\\
95	0.709886195024686\\
96	0.708040510988342\\
97	0.706218419715735\\
98	0.704418150628316\\
99	0.702637837472527\\
100	0.700875532073147\\
};
\addlegendentry{Speed = 130km/h}

\end{axis}
\end{tikzpicture}%

%% file: Figures/T_T.tex
\definecolor{mycolor1}{rgb}{0.00000,0.75000,0.75000}%
\begin{tikzpicture}
\pgfplotsset{
tick label style={font=\footnotesize},
label style={font=\footnotesize},
legend style={font=\scriptsize}
}
\tikzset{every mark/.append style={scale=1.2}}
\begin{axis}[%
width=\fwidth,
height=\fheight,
at={(0\fwidth,0\fheight)},
scale only axis,
xmin=0,
xmax=100,
xlabel style={font=\color{white!15!black}},
xlabel={$\text{Nodes per km (}\rho\text{)}$},
xlabel style={font=\color{white!15!black}},
ymin=0,
ymax=1,
xmajorgrids,
ymajorgrids,
ylabel style={font=\color{white!15!black}},
ylabel={Communication Duration [Ratio]},
axis background/.style={fill=white},
legend columns = 1,
legend pos =south west,
legend style={legend cell align=left, align=left, draw=white!15!black},
cycle list={%
{blue,mark=*, mark repeat = 5},
{red,mark=square, mark repeat = 5},
{black,mark=o, mark repeat = 5},
{brown!60!black,
mark options={fill=brown!40, scale = 1.2, mark repeat = 5},
mark=triangle*}}
]
\addplot 
  table[row sep=crcr]{%
1	0.276539473183165\\
2	0.469169465939936\\
3	0.605030700468574\\
4	0.702024566068947\\
5	0.77208956431113\\
6	0.823277879977911\\
7	0.861082053938158\\
8	0.889290530613844\\
9	0.910544948246689\\
10	0.926706911729394\\
11	0.939101891676616\\
12	0.948683086455335\\
13	0.956142626944714\\
14	0.961987792482161\\
15	0.96659375018603\\
16	0.970240390949604\\
17	0.973138291553548\\
18	0.97544717409062\\
19	0.977289142947207\\
20	0.978758255356524\\
21	0.97992749648925\\
22	0.980853902381159\\
23	0.98158235078054\\
24	0.982148386691116\\
25	0.982580343253328\\
26	0.98290094456403\\
27	0.983128524990109\\
28	0.983277962682788\\
29	0.983361398723693\\
30	0.983388794469371\\
31	0.983368366025342\\
32	0.983306924862272\\
33	0.983210146324484\\
34	0.983082782433362\\
35	0.982928831424076\\
36	0.982751673501487\\
37	0.982554180086465\\
38	0.982338802156673\\
39	0.982107642021171\\
40	0.981862511906467\\
41	0.981604981994213\\
42	0.981336419984708\\
43	0.981058023821662\\
44	0.980770848874067\\
45	0.980475830605045\\
46	0.980173803549924\\
47	0.979865517261688\\
48	0.979551649752581\\
49	0.979232818857816\\
50	0.978909591865116\\
51	0.978582493688497\\
52	0.97825201381157\\
53	0.977918612183464\\
54	0.977582724215702\\
55	0.97724476500089\\
56	0.976905132850975\\
57	0.976564212234958\\
58	0.976222376180556\\
59	0.975879988192759\\
60	0.975537403732167\\
61	0.975194971288769\\
62	0.974853033080571\\
63	0.974511925402388\\
64	0.97417197864664\\
65	0.973833517016301\\
66	0.973496857948817\\
67	0.973162311269973\\
68	0.972830178096995\\
69	0.972500749511862\\
70	0.972174305027274\\
71	0.971851110870576\\
72	0.971531418113268\\
73	0.971215460677085\\
74	0.970903453250589\\
75	0.970595589153609\\
76	0.970292038189666\\
77	0.969992944529593\\
78	0.969698424671629\\
79	0.969408565525234\\
80	0.969123422666724\\
81	0.968843018815012\\
82	0.968567342574748\\
83	0.968296347492071\\
84	0.968029951464876\\
85	0.967768036544785\\
86	0.967510449162083\\
87	0.967257000797675\\
88	0.967007469117497\\
89	0.966761599575511\\
90	0.966519107480799\\
91	0.966279680513455\\
92	0.966042981662499\\
93	0.96580865254788\\
94	0.965576317077822\\
95	0.965345585382624\\
96	0.965116057957573\\
97	0.964887329940253\\
98	0.964658995442782\\
99	0.96443065185652\\
100	0.964201904046869\\
};
\addlegendentry{$\text{T}_{\text{RT}}=\text{ 0.025s}$}

\addplot
  table[row sep=crcr]{%
1	0.273399519893544\\
2	0.464011990342117\\
3	0.598289264798796\\
4	0.693888503094023\\
5	0.762662337536757\\
6	0.812631254607136\\
7	0.849273446089094\\
8	0.876368089356224\\
9	0.896549013858502\\
10	0.911670679171597\\
11	0.923052063873606\\
12	0.931640639828744\\
13	0.938123636001846\\
14	0.943004241401007\\
15	0.946654280942453\\
16	0.949350962137391\\
17	0.951302740118282\\
18	0.952667683206801\\
19	0.953566624979632\\
20	0.954092661130224\\
21	0.954318062373633\\
22	0.954299345920133\\
23	0.954081024357189\\
24	0.95369839733802\\
25	0.953179645388068\\
26	0.952547411228363\\
27	0.951820002132255\\
28	0.951012310146551\\
29	0.950136520884006\\
30	0.949202662861742\\
31	0.948219035837874\\
32	0.947192546773169\\
33	0.946128974859947\\
34	0.945033181774208\\
35	0.943909279394111\\
36	0.942760764314638\\
37	0.941590626306742\\
38	0.940401436226622\\
39	0.939195417637198\\
40	0.937974505457313\\
41	0.936740394230046\\
42	0.935494578044722\\
43	0.934238383716864\\
44	0.932972998496187\\
45	0.931699493311962\\
46	0.930418842360619\\
47	0.929131939679435\\
48	0.927839613222666\\
49	0.926542636855177\\
50	0.925241740597732\\
51	0.923937619393149\\
52	0.922630940610369\\
53	0.921322350461142\\
54	0.920012479469875\\
55	0.918701947109258\\
56	0.917391365691654\\
57	0.916081343587769\\
58	0.914772487829051\\
59	0.913465406138244\\
60	0.912160708422704\\
61	0.910859007757463\\
62	0.909560920879173\\
63	0.908267068207679\\
64	0.906978073409074\\
65	0.905694562512358\\
66	0.904417162591262\\
67	0.903146500023272\\
68	0.901883198339253\\
69	0.900627875679393\\
70	0.89938114187418\\
71	0.898143595172929\\
72	0.89691581864666\\
73	0.895698376297023\\
74	0.894491808908154\\
75	0.893296629683849\\
76	0.892113319718004\\
77	0.890942323351832\\
78	0.889784043476686\\
79	0.888638836846235\\
80	0.887507009466142\\
81	0.88638881213295\\
82	0.885284436196588\\
83	0.884194009622437\\
84	0.883117593429207\\
85	0.882055178577758\\
86	0.8810066833834\\
87	0.879971951520029\\
88	0.878950750678685\\
89	0.87794277193579\\
90	0.876947629877475\\
91	0.875964863516194\\
92	0.874993938024402\\
93	0.874034247297679\\
94	0.873085117346553\\
95	0.872145810502736\\
96	0.871215530411853\\
97	0.870293427771322\\
98	0.869378606759187\\
99	0.868470132087727\\
100	0.867567036604799\\
};
\addlegendentry{$\text{T}_{\text{RT}}=\text{ 0.1s}$}

\addplot
  table[row sep=crcr]{%
1	0.269226270010446\\
2	0.457174065072578\\
3	0.58936908684978\\
4	0.683142004614403\\
5	0.750231264366396\\
6	0.798615295332723\\
7	0.833753150199068\\
8	0.859411529347692\\
9	0.878213705695854\\
10	0.892004490814542\\
11	0.902094126841523\\
12	0.909422410303815\\
13	0.914670027495374\\
14	0.91833473114857\\
15	0.920783923434728\\
16	0.922291276876314\\
17	0.923062464899221\\
18	0.923253398707182\\
19	0.922983263754648\\
20	0.92234391687206\\
21	0.921406715418205\\
22	0.920227519758967\\
23	0.918850386106493\\
24	0.91731031317266\\
25	0.915635300095524\\
26	0.913847899382712\\
27	0.91196639696591\\
28	0.910005715005823\\
29	0.907978107172994\\
30	0.905893697580382\\
31	0.90376090117568\\
32	0.901586753703742\\
33	0.899377172269223\\
34	0.897137162327705\\
35	0.894870983088467\\
36	0.892582280452829\\
37	0.890274194473478\\
38	0.88794944671165\\
39	0.885610411652373\\
40	0.883259175412675\\
41	0.880897584270305\\
42	0.878527284996745\\
43	0.876149758558279\\
44	0.873766348422735\\
45	0.871378284454932\\
46	0.868986703184274\\
47	0.866592665070513\\
48	0.864197169269091\\
49	0.861801166298236\\
50	0.859405568930638\\
51	0.857011261568778\\
52	0.854619108311518\\
53	0.852229959877884\\
54	0.849844659520073\\
55	0.847464048030077\\
56	0.845088967921771\\
57	0.842720266851924\\
58	0.840358800328617\\
59	0.838005433743472\\
60	0.835661043754458\\
61	0.833326519038495\\
62	0.831002760427411\\
63	0.828690680436718\\
64	0.826391202194178\\
65	0.824105257773892\\
66	0.821833785941755\\
67	0.819577729319371\\
68	0.817338030975922\\
69	0.815115630460878\\
70	0.812911459294877\\
71	0.810726435941351\\
72	0.808561460287625\\
73	0.806417407671012\\
74	0.80429512249283\\
75	0.802195411471195\\
76	0.800119036591601\\
77	0.79806670782267\\
78	0.796039075672785\\
79	0.79403672367138\\
80	0.792060160866323\\
81	0.790109814435786\\
82	0.788186022519061\\
83	0.786289027375727\\
84	0.784418968986163\\
85	0.782575879208412\\
86	0.780759676606695\\
87	0.778970162065185\\
88	0.777207015296981\\
89	0.775469792352323\\
90	0.773757924222056\\
91	0.772070716622078\\
92	0.770407351032057\\
93	0.768766887047283\\
94	0.767148266086085\\
95	0.765550316477205\\
96	0.763971759932002\\
97	0.76241121938564\\
98	0.76086722816993\\
99	0.759338240458453\\
100	0.75782264290244\\
};
\addlegendentry{$\text{T}_{\text{RT}}=\text{ 0.2s}$}

\addplot 
  table[row sep=crcr]{%
1	0.256797539662996\\
2	0.436922794937109\\
3	0.56307051030683\\
4	0.651586096983991\\
5	0.713867355241906\\
6	0.757767675310445\\
7	0.788688354172194\\
8	0.810357573884616\\
9	0.825365719743624\\
10	0.835527627365945\\
11	0.84212658613976\\
12	0.846078902088636\\
13	0.848045393426718\\
14	0.848507424538926\\
15	0.84781915298935\\
16	0.846243727470563\\
17	0.843978581345728\\
18	0.841173260205861\\
19	0.837942096460419\\
20	0.834373298186841\\
21	0.830535522252432\\
22	0.826482667972285\\
23	0.822257401910791\\
24	0.817893770718809\\
25	0.813419153395021\\
26	0.80885573139882\\
27	0.804221604207402\\
28	0.799531642229444\\
29	0.794798143762325\\
30	0.790031344718752\\
31	0.785239816970764\\
32	0.780430781862638\\
33	0.775610358688444\\
34	0.770783762988208\\
35	0.765955465878756\\
36	0.761129322940351\\
37	0.756308679171476\\
38	0.751496455017797\\
39	0.746695217344998\\
40	0.741907238362942\\
41	0.737134544850508\\
42	0.73237895952532\\
43	0.727642136012602\\
44	0.722925588564654\\
45	0.718230717446053\\
46	0.713558830714125\\
47	0.708911162977685\\
48	0.704288891600764\\
49	0.699693150725159\\
50	0.695125043411159\\
51	0.69058565213571\\
52	0.686076047838598\\
53	0.681597297667613\\
54	0.67715047154136\\
55	0.67273664762194\\
56	0.668356916768151\\
57	0.664012386022244\\
58	0.65970418116902\\
59	0.655433448394679\\
60	0.651201355063872\\
61	0.647009089626708\\
62	0.642857860662658\\
63	0.638748895065312\\
64	0.634683435370609\\
65	0.630662736231402\\
66	0.626688060042902\\
67	0.622760671726711\\
68	0.618881832685583\\
69	0.6150527939469\\
70	0.611274788519808\\
71	0.607549022999164\\
72	0.603876668458705\\
73	0.600258850686028\\
74	0.596696639823077\\
75	0.593191039487576\\
76	0.58974297546311\\
77	0.586353284058202\\
78	0.583022700247417\\
79	0.579751845720081\\
80	0.576541216974381\\
81	0.573391173605969\\
82	0.570301926950648\\
83	0.567273529249687\\
84	0.564305863513694\\
85	0.561398634266239\\
86	0.558551359351372\\
87	0.555763362989429\\
88	0.553033770262783\\
89	0.55036150320724\\
90	0.54774527867531\\
91	0.545183608124532\\
92	0.542674799467156\\
93	0.540216961096876\\
94	0.537808008183849\\
95	0.535445671301172\\
96	0.533127507414466\\
97	0.530850913231441\\
98	0.528613140870926\\
99	0.526411315770993\\
100	0.52424245671442\\
};
\addlegendentry{$\text{T}_{\text{RT}}=\text{ 0.5s}$}

\addplot [color=red!30!yellow, mark=diamond, mark options={ scale = 1.2}, mark repeat = 5]
  table[row sep=crcr]{%
1	0.236382919667243\\
2	0.404027551912086\\
3	0.520735750717365\\
4	0.601192003232198\\
5	0.656231740968311\\
6	0.693502147656798\\
7	0.718305586045371\\
8	0.73430110967333\\
9	0.744019154741332\\
10	0.749220236197538\\
11	0.751139403998056\\
12	0.750651452470347\\
13	0.748382469960257\\
14	0.744785407800297\\
15	0.740191572748544\\
16	0.734845970943522\\
17	0.728931767304808\\
18	0.722587360682464\\
19	0.715918412118649\\
20	0.709006396150236\\
21	0.701914736831711\\
22	0.694693251777555\\
23	0.687381400810715\\
24	0.680010682850821\\
25	0.672606420730669\\
26	0.665189102460128\\
27	0.657775398359752\\
28	0.650378939360138\\
29	0.643010917864225\\
30	0.635680555708447\\
31	0.628395471774743\\
32	0.62116197322605\\
33	0.613985288151503\\
34	0.606869752914931\\
35	0.599818964214623\\
36	0.592835903442665\\
37	0.585923039137859\\
38	0.579082411986501\\
39	0.572315705818213\\
40	0.565624307281752\\
41	0.559009356304928\\
42	0.5524717889972\\
43	0.546012374309457\\
44	0.539631745497975\\
45	0.533330427230221\\
46	0.527108859005246\\
47	0.52096741543066\\
48	0.514906423793859\\
49	0.508926179281374\\
50	0.50302695813261\\
51	0.497209028959281\\
52	0.491472662417064\\
53	0.485818139379301\\
54	0.480245757732466\\
55	0.474755837888347\\
56	0.469348727087622\\
57	0.464024802552915\\
58	0.458784473536028\\
59	0.453628182293399\\
60	0.448556404015561\\
61	0.44356964573034\\
62	0.43866844419534\\
63	0.433853362793004\\
64	0.429124987440861\\
65	0.424483921530588\\
66	0.419930779911972\\
67	0.415466181941855\\
68	0.411090743623469\\
69	0.406805068868279\\
70	0.402609739920381\\
71	0.398505306992672\\
72	0.394492277174187\\
73	0.390571102679236\\
74	0.386742168520958\\
75	0.383005779704674\\
76	0.379362148049549\\
77	0.375811378760629\\
78	0.372353456886764\\
79	0.368988233813257\\
80	0.365715413950812\\
81	0.362534541794262\\
82	0.359444989535272\\
83	0.356445945422362\\
84	0.353536403068822\\
85	0.350715151913992\\
86	0.347980769045627\\
87	0.345331612590185\\
88	0.342765816873647\\
89	0.340281289547463\\
90	0.337875710862212\\
91	0.335546535255301\\
92	0.333290995398356\\
93	0.331106108824802\\
94	0.328988687228495\\
95	0.326935348490231\\
96	0.324942531450833\\
97	0.323006513407487\\
98	0.32112343026477\\
99	0.319289299223716\\
100	0.317500043842372\\
};
\addlegendentry{$\text{T}_{\text{RT}}=\text{ 1s}$}

\end{axis}
\end{tikzpicture}%

%% file: Figures/B_MIMO.tex
\usetikzlibrary{plotmarks}
\begin{tikzpicture}

\pgfplotsset{
tick label style={font=\footnotesize},
label style={font=\footnotesize},
legend style={font=\scriptsize}
}

\tikzset{every mark/.append style={scale=1.5}}

\begin{axis}[%
width=\fwidth,
height=\fheight,
at={(0\fwidth,0\fheight)},
scale only axis,
xmin=0,
xmax=84,
xlabel={Nodes per km ($\rho$)},
ylabel={Throughput [Gbps]},
ymin=0,
ymax=5,
axis background/.style={fill=white},
xmajorgrids,
ymajorgrids,
legend columns={2},
legend style={at={(1,1.2)},legend cell align=left, align=center, draw=white!15!black},
legend pos = south east,
cycle list={%
{blue,mark=*, mark repeat=2},
{red,mark=square, mark repeat=2},
{black,mark=o,, mark repeat=2},
{brown!60!black,
mark options={fill=brown!40, scale = 2},
mark=triangle*}}
]
\addplot 
  table[row sep=crcr]{%
5	0\\
6	5.61886275047415e-14\\
7	0.0290716404140761\\
8	0.0498488818759088\\
9	0.0815086822776651\\
10	0.131563565139342\\
11	0.176066962294916\\
12	0.246691249915429\\
13	0.311204081129938\\
14	0.39957256104033\\
15	0.47611867916927\\
16	0.529630717820306\\
17	0.66788857575951\\
18	0.799359351404522\\
19	0.824855914466146\\
20	1.00243554572634\\
21	1.07418929032967\\
22	1.13762253316493\\
23	1.20949455446347\\
24	1.29700304009932\\
25	1.39605558312932\\
27	1.41580269413093\\
28	1.54651885604961\\
30	1.55031812887739\\
32	1.64334472309335\\
34	1.69785830984043\\
36	1.80133203150852\\
39	1.80674904952623\\
42	1.88417555909496\\
46	1.88434510466225\\
50	1.89104030093251\\
56	1.80316058975549\\
63	1.77743355551019\\
72	1.74420099245265\\
84	1.76284392640831\\
};
\addlegendentry{4x4}

\addplot 
  table[row sep=crcr]{%
5	0\\
6	1.69155491997151e-13\\
7	0.0834927524420294\\
8	0.134844824646133\\
9	0.21093525145167\\
10	0.323152024407122\\
11	0.411312711103312\\
12	0.546699475150452\\
13	0.656234365911404\\
14	0.800271866002058\\
15	0.91644366848481\\
16	0.987128901288949\\
17	1.18278349416223\\
18	1.35683692695309\\
19	1.37642407571198\\
20	1.59521154133697\\
21	1.6739833368881\\
22	1.731766055046\\
23	1.79682646488403\\
24	1.89403234062927\\
25	1.99294961030942\\
27	1.97232067586097\\
28	2.10112914512188\\
30	2.06927856170497\\
32	2.14808327895365\\
34	2.16537281444541\\
36	2.2528164756446\\
39	2.21619023500481\\
42	2.25446132099127\\
46	2.21083330788088\\
50	2.16838978604522\\
56	2.02500627885266\\
63	1.9538037305088\\
72	1.90439831777904\\
84	1.92905297297505\\
};
\addlegendentry{4x16}

\addplot 
  table[row sep=crcr]{%
5	0\\
6	5.03639900269429e-13\\
7	0.232627677176137\\
8	0.368062483908449\\
9	0.550556115766518\\
10	0.811461460508262\\
11	0.985580230395391\\
12	1.24714035082353\\
13	1.46374834342802\\
14	1.71605788652696\\
15	1.88767203453792\\
16	2.02789796966826\\
17	2.3012842561857\\
18	2.5574704788173\\
19	2.58214182653544\\
20	2.88540311235165\\
21	3.00841753053498\\
22	3.09554158378638\\
23	3.18777104155679\\
24	3.29548943095973\\
25	3.42272218077195\\
27	3.42905621144381\\
28	3.6107167517368\\
30	3.57974361703247\\
32	3.69074689510015\\
34	3.71595160269559\\
36	3.82504301786151\\
39	3.78167698398203\\
42	3.87530963905723\\
46	3.82461639615576\\
50	3.78136988546501\\
56	3.65883983691273\\
63	3.56859048749613\\
72	3.49991068332567\\
84	3.50738781604701\\
};
\addlegendentry{64x4}

\addplot
  table[row sep=crcr]{%
5	0\\
6	1.5200235875876e-12\\
7	0.592278571750952\\
8	0.861316019722894\\
9	1.17774675492432\\
10	1.5820672850151\\
11	1.82012724633143\\
12	2.15895845867519\\
13	2.41433318602194\\
14	2.69406284214404\\
15	2.88240905547986\\
16	3.02260682726115\\
17	3.31738246936676\\
18	3.56805842101553\\
19	3.56075438912099\\
20	3.86656382570524\\
21	3.96512130758042\\
22	4.0394499550445\\
23	4.11848051951412\\
24	4.20289031968821\\
25	4.30884482113473\\
27	4.25759480383992\\
28	4.42589712172106\\
30	4.34412056402745\\
32	4.42557145734503\\
34	4.40191307624979\\
36	4.47765203988995\\
39	4.37193351909432\\
42	4.42019748545825\\
46	4.31593979623955\\
50	4.21184862166414\\
56	4.01190152981434\\
63	3.86259311698214\\
72	3.76974148584263\\
84	3.8008998975864\\
};
\addlegendentry{64x16}

\end{axis}
\end{tikzpicture}%

%% file: Figures/B_V.tex
\definecolor{mycolor1}{rgb}{0.00000,0.75000,0.75000}%
\begin{tikzpicture}

\pgfplotsset{
tick label style={font=\footnotesize},
label style={font=\footnotesize},
legend style={font=\scriptsize}
}

\tikzset{every mark/.append style={scale=1.5}}

\begin{axis}[%
width=\fwidth,
height=\fheight,
at={(0\fwidth,0\fheight)},
scale only axis,
xmin=0,
xmax=84,
xlabel={Nodes per km ($\rho$)},
ylabel={Throughput [Gbps]},
ymin=0,
ymax=5,
axis background/.style={fill=white},
xmajorgrids,
ymajorgrids,
legend columns={1},
legend pos = south east,
legend cell align = left,
cycle list={%
{blue,mark=*, mark repeat=2},
{red,mark=square, mark repeat=2},
{black,mark=o,, mark repeat=2},
{brown!60!black,
mark options={fill=brown!40, scale = 2},
mark=triangle*}}
]
\addplot 
  table[row sep=crcr]{%
5	0\\
6	1.84750383260615e-12\\
7	0.691327817518567\\
8	0.977485705935555\\
9	1.31022488251016\\
10	1.73521178769099\\
11	1.9755069781237\\
12	2.32592608847518\\
13	2.58699960972768\\
14	2.8758109335849\\
15	3.0685526821679\\
16	3.21206113205919\\
17	3.52132572446825\\
18	3.78476606372141\\
19	3.77636904893884\\
20	4.09969045326838\\
21	4.20529323043752\\
22	4.28578836316651\\
23	4.3715550523505\\
24	4.46300296052292\\
25	4.57695743882206\\
27	4.53128654859735\\
28	4.71014647512764\\
30	4.63041602219934\\
32	4.72289056425255\\
34	4.70043319771115\\
36	4.77998355120204\\
39	4.67126874466033\\
42	4.71969090196802\\
46	4.60666779421827\\
50	4.48057102570346\\
56	4.25730879305212\\
63	4.07451235675762\\
72	3.93683675827021\\
84	3.90050707402412\\
100	4.08871829352067\\
};
\addlegendentry{Speed = 10km/h}

\addplot 
  table[row sep=crcr]{%
5	0\\
6	1.80230123278478e-12\\
7	0.676608389805946\\
8	0.959151545981971\\
9	1.28828229410458\\
10	1.70914314422398\\
11	1.94808590487178\\
12	2.2962783305551\\
13	2.55620458857143\\
14	2.84397984675046\\
15	3.03628194831818\\
16	3.18004033047514\\
17	3.48813391965935\\
18	3.75112682117333\\
19	3.74381227239334\\
20	4.06651637651096\\
21	4.17236781231229\\
22	4.25334996347935\\
23	4.33959965642354\\
24	4.43152704680571\\
25	4.54584675304618\\
27	4.50163502698235\\
28	4.68050954175118\\
30	4.6024366080816\\
32	4.69552187720163\\
34	4.67434911528282\\
36	4.75462229392244\\
39	4.64761245479854\\
42	4.69691923455593\\
46	4.58553385495569\\
50	4.46106673409379\\
56	4.23976194874023\\
63	4.05864509392807\\
72	3.92237610041918\\
84	3.88700400489394\\
100	4.07536140104183\\
};
\addlegendentry{Speed = 20km/h}

\addplot 
  table[row sep=crcr]{%
5	0\\
6	1.758518419328e-12\\
7	0.6622886686618\\
8	0.941254955038922\\
9	1.26680665833301\\
10	1.68357132952423\\
11	1.92114683768985\\
12	2.26710790282928\\
13	2.52587131126683\\
14	2.81259091823814\\
15	3.00443579463991\\
16	3.14841736271448\\
17	3.4553302280923\\
18	3.71785637410632\\
19	3.7116005605448\\
20	4.03366976105909\\
21	4.13975543187941\\
22	4.22120818626924\\
23	4.30792486227247\\
24	4.4003160978574\\
25	4.51498666421351\\
27	4.47221158291169\\
28	4.65108980762886\\
30	4.57465207288282\\
32	4.66833385154846\\
34	4.64842770189995\\
36	4.72940992844114\\
39	4.62408637787118\\
42	4.67426454037866\\
46	4.56450068246687\\
50	4.441648222766\\
56	4.22228576453546\\
63	4.04283582586005\\
72	3.90796290854585\\
84	3.87354021752091\\
100	4.06203834849603\\
};
\addlegendentry{Speed = 30km/h}

\addplot 
  table[row sep=crcr]{%
5	0\\
6	1.48781139254868e-12\\
7	0.57226640173281\\
8	0.827286357364481\\
9	1.12864921725865\\
10	1.51761640168319\\
11	1.74529499381888\\
12	2.07557588316532\\
13	2.32583158494868\\
14	2.6046840155804\\
15	2.79288667778128\\
16	2.93774040482678\\
17	3.23615290419262\\
18	3.49491620069117\\
19	3.49544256741692\\
20	3.81261358725802\\
21	3.91995864061145\\
22	4.00427032406662\\
23	4.09383023026697\\
24	4.18905211785959\\
25	4.3057959990717\\
27	4.27247133383611\\
28	4.45108599126633\\
30	4.38549139026532\\
32	4.4829659186422\\
34	4.47143877205704\\
36	4.55701158071742\\
39	4.46298361799339\\
42	4.51890148015253\\
46	4.42004563010355\\
50	4.30808570977853\\
56	4.10190472865691\\
63	3.93377525636159\\
72	3.80838523024432\\
84	3.78038302571375\\
100	3.96971654982133\\
};
\addlegendentry{Speed = 100km/h}

\addplot [color=red!30!yellow, mark=diamond, mark options={ scale = 2}]
  table[row sep=crcr]{%
5	0\\
6	1.38863015633521e-12\\
7	0.538578634453808\\
8	0.783932810946686\\
9	1.07541212088315\\
10	1.4529548364664\\
11	1.67626897877675\\
12	1.99983817912505\\
13	2.24629156256702\\
14	2.52155899167535\\
15	2.70799475011683\\
16	2.85288848210029\\
17	3.1475543124108\\
18	3.40446732539205\\
19	3.40758545997042\\
20	3.72243823441851\\
21	3.83013399687542\\
22	3.91545293286943\\
23	4.00601767847835\\
24	4.10224321962337\\
25	4.21968293611999\\
27	4.19009917144384\\
28	4.36845522984377\\
30	4.3071982559451\\
32	4.40610283109021\\
34	4.39791611790993\\
36	4.4852648937199\\
39	4.39581481494589\\
42	4.45400654709945\\
46	4.35959555084523\\
50	4.25209028015518\\
56	4.05134177424914\\
63	3.8878818286089\\
72	3.76640405927273\\
84	3.74103523050595\\
100	3.93064824812382\\
};
\addlegendentry{Speed = 130km/h}

\end{axis}
\end{tikzpicture}%

%% file: Figures/B_T.tex
\definecolor{mycolor1}{rgb}{0.00000,0.75000,0.75000}%
\begin{tikzpicture}

\pgfplotsset{
tick label style={font=\footnotesize},
label style={font=\footnotesize},
legend style={font=\scriptsize}
}

\tikzset{every mark/.append style={scale=1.5}}

\begin{axis}[%
width=\fwidth,
height=\fheight,
at={(0\fwidth,0\fheight)},
scale only axis,
xmin=0,
xmax=84,
xlabel={Nodes per km ($\rho$)},
ylabel={Throughput [Gbps]},
ymin=0,
ymax=5,
axis background/.style={fill=white},
xmajorgrids,
ymajorgrids,
legend columns={1},
legend pos = south east,
legend cell align = left,
cycle list={%
{blue,mark=*, mark repeat=2},
{red,mark=square, mark repeat=2},
{black,mark=o,, mark repeat=2},
{brown!60!black,
mark options={fill=brown!40, scale = 2},
mark=triangle*}}
]
\addplot 
  table[row sep=crcr]{%
5	0\\
6	1.83606726387464e-12\\
7	0.687609805958696\\
8	0.972860489452497\\
9	1.30469483732543\\
10	1.72864745829846\\
11	1.96860599378507\\
12	2.31846892108718\\
13	2.57925713561838\\
14	2.86781132822353\\
15	3.06044484779839\\
16	3.2040183284049\\
17	3.51299110665262\\
18	3.77632142842168\\
19	3.76819727889677\\
20	4.09136603500265\\
21	4.19703234498332\\
22	4.27765078764687\\
23	4.36353974514375\\
24	4.455109004338\\
25	4.56915614976306\\
27	4.52385217812696\\
28	4.70271678164937\\
30	4.62340281555218\\
32	4.7160313824593\\
34	4.69389686484487\\
36	4.77362922489836\\
39	4.66534242103063\\
42	4.71398698229408\\
46	4.60137483331207\\
50	4.47568688799846\\
56	4.25291544033504\\
63	4.07054009119549\\
72	3.93321713442828\\
84	3.89712761714316\\
100	4.08537589269722\\
};
\addlegendentry{$\text{T}_{\text{RTO}}$ = $\text{ 0.025s}$}

\addplot 
  table[row sep=crcr]{%
5	0\\
6	1.67501356853001e-12\\
7	0.634799966859139\\
8	0.906727892698954\\
9	1.22521188432267\\
10	1.63387589261287\\
11	1.86867686441733\\
12	2.2101646905921\\
13	2.46655904000043\\
14	2.75111207908017\\
15	2.94199225522032\\
16	3.08634283320524\\
17	3.39086690004799\\
18	3.65240379810872\\
19	3.64819598950235\\
20	3.96894447682958\\
21	4.0754564943813\\
22	4.15780239128702\\
23	4.24540608400314\\
24	4.33867914505428\\
25	4.45400927776492\\
27	4.41404111007945\\
28	4.59289484879418\\
30	4.51966159926135\\
32	4.61449448216222\\
34	4.59706837598791\\
36	4.67942799392481\\
39	4.57742168571257\\
42	4.62930341125628\\
46	4.52273451282468\\
50	4.40306687399831\\
56	4.18754412003278\\
63	4.01139033821324\\
72	3.87927823451361\\
84	3.84672998255235\\
100	4.03549338323295\\
};
\addlegendentry{$\text{T}_{\text{RTO}}$ = $\text{ 0.1s}$}

\addplot 
  table[row sep=crcr]{%
5	0\\
6	1.48781139254868e-12\\
7	0.57226640173281\\
8	0.827286357364481\\
9	1.12864921725865\\
10	1.51761640168319\\
11	1.74529499381888\\
12	2.07557588316532\\
13	2.32583158494868\\
14	2.6046840155804\\
15	2.79288667778128\\
16	2.93774040482678\\
17	3.23615290419262\\
18	3.49491620069117\\
19	3.49544256741692\\
20	3.81261358725802\\
21	3.91995864061145\\
22	4.00427032406662\\
23	4.09383023026697\\
24	4.18905211785959\\
25	4.3057959990717\\
27	4.27247133383611\\
28	4.45108599126633\\
30	4.38549139026532\\
32	4.4829659186422\\
34	4.47143877205704\\
36	4.55701158071742\\
39	4.46298361799339\\
42	4.51890148015253\\
46	4.42004563010355\\
50	4.30808570977853\\
56	4.10190472865691\\
63	3.93377525636159\\
72	3.80838523024432\\
84	3.78038302571375\\
100	3.96971654982133\\
};
\addlegendentry{$\text{T}_{\text{RTO}}$ = $\text{ 0.2s}$}

\addplot 
  table[row sep=crcr]{%
5	0\\
6	1.07032176147248e-12\\
7	0.42739896114555\\
8	0.637666452808048\\
9	0.892616682495383\\
10	1.22751493036538\\
11	1.43312970207385\\
12	1.73029128598334\\
13	1.96100598861615\\
14	2.22108529898946\\
15	2.39953464586673\\
16	2.54296442536444\\
17	2.82225413141031\\
18	3.07063575900182\\
19	3.08247150151077\\
20	3.38699580067111\\
21	3.49512014374137\\
22	3.58332647294976\\
23	3.67678607549453\\
24	3.77591841933783\\
25	3.89512105337981\\
27	3.87881744678454\\
28	4.05537041073771\\
30	4.00976200723765\\
32	4.11332285605476\\
34	4.11711296851054\\
36	4.21051064366533\\
39	4.13789955328709\\
42	4.20414887477523\\
46	4.12621901230378\\
50	4.03532119921666\\
56	3.85506571661293\\
63	3.7092411075426\\
72	3.60254000668895\\
84	3.58702378218876\\
100	3.77730333175903\\
};
\addlegendentry{$\text{T}_{\text{RTO}} $ = $\text{ 0.5s}$}

\addplot [color=red!30!yellow, mark=diamond, mark options={ scale = 2}]
  table[row sep=crcr]{%
5	0\\
6	6.70554268781015e-13\\
7	0.279205338193522\\
8	0.433118142400496\\
9	0.626833497485468\\
10	0.888290685736135\\
11	1.05865883729679\\
12	1.30524929726689\\
13	1.50304233965657\\
14	1.72998079322826\\
15	1.88925230080416\\
16	2.02400115040701\\
17	2.27086189250382\\
18	2.4978083700755\\
19	2.52116320828896\\
20	2.80066569985582\\
21	2.90590186730783\\
22	2.99554855306192\\
23	3.09049248563478\\
24	3.1911612599298\\
25	3.30986831269295\\
27	3.31397221732883\\
28	3.48365733947317\\
30	3.46317189915759\\
32	3.57185362779175\\
34	3.59446034286929\\
36	3.69581549447705\\
39	3.65161052950645\\
42	3.72997181641222\\
46	3.68040356483049\\
50	3.61848513799756\\
56	3.47511735384112\\
63	3.36110866251036\\
72	3.28104833432297\\
84	3.28281184021842\\
100	3.47233463382795\\
};
\addlegendentry{$\text{T}_{\text{RTO}}$ = $\text{ 1s}$}

\end{axis}
\end{tikzpicture}%

%% file: Figures/Hist_B.tex
\definecolor{ref}{rgb}{0.65,0.65,0.65} %{0.4,0.8,0.85}
\definecolor{lhmm}{rgb}{0.9,0.6,0.5}
\definecolor{ghmm}{rgb}{0.7,0.9,0.35}
\definecolor{lsvm}{rgb}{0.9,0.8,0.25}
\definecolor{gsvm}{rgb}{0.4,0.8,0.9}
\usetikzlibrary{patterns}
\begin{tikzpicture}

\begin{axis}[%
width=0.993\fwidth,
height=\fheight,
at={(0\fwidth,0\fheight)},
scale only axis,
bar shift auto,
xmin=0,
xmax=6,
xtick={1,2,3,4,5},
xticklabels={{10},{20},{30},{100},{130}},
xlabel style={font=\color{white!15!black}},
xlabel={Speed [m/s]},
ymin=0,
ymax=5,
ybar,
ylabel style={font=\color{white!15!black}},
ylabel={Throughput [Gbps]},
axis background/.style={fill=white},
title style={font=\bfseries},
xmajorgrids,
ymajorgrids,
legend style={legend cell align=left, align=left, draw=white!15!black, at={(0.5,1.15)},/tikz/every even column/.append style={column sep=0.7cm},
  anchor=north,legend columns=-1},
]
\addplot[ bar width=0.145,fill=lhmm, draw=black, postaction={pattern=crosshatch}]  table[row sep=crcr] {%
1	1.07685147720872\\
2	1.06725832462293\\
3	1.05774999799827\\
4	0.993498258717012\\
5	0.967153922499262\\
};
\addlegendentry{4x4}

\addplot[bar width=0.145,fill=ghmm, draw=black,postaction={pattern=crosshatch dots}]  table[row sep=crcr] {
1	1.72\\
2	1.69017207169064\\
3	1.67625250659605\\
4	1.5821070525637\\
5	1.54346163897122\\
};
\addlegendentry{4x16}

\addplot[bar width=0.145,fill=lsvm,draw=black, postaction={pattern=north east lines}]  table[row sep=crcr] {%
1	3.05694542140489\\
2	3.03487818107832\\
3	3.01299264854539\\
4	2.86474376522053\\
5	2.80377248878629\\
};
\addlegendentry{64x4}

\addplot[ bar width=0.145,fill=gsvm,draw=black] table[row sep=crcr] {%
1	4.08240959103413\\
2	4.05466046649173\\
3	4.02713486147692\\
4	3.84054759490414\\
5	3.76373919132898\\
};
\addlegendentry{64x16}

\end{axis}
\end{tikzpicture}%